\documentclass[aps,prx,twocolumn,showpacs,superscriptaddress,groupedaddress]{revtex4-1}  
\usepackage{amssymb,amsmath,graphicx}
\usepackage{graphicx,wrapfig,lipsum}
\usepackage[usenames,dvipsnames]{xcolor}
\usepackage{graphicx}  
\usepackage{dcolumn}   
\usepackage{bm}        
\usepackage{amssymb}   
\usepackage{braket}
\usepackage{mathtools}
\usepackage{hyperref}
\usepackage[dvipsnames]{xcolor}
\hypersetup{
    colorlinks,
    linkcolor={blue},
    citecolor={blue},
    urlcolor={blue}
}
\usepackage{mathtools}
\usepackage{stackengine}
\newcommand\barbelow[1]{\stackunder[1.2pt]{$#1$}{\rule{.8ex}{.075ex}}}

\newcommand{\vect}[1]{\mathbf{#1}}

\usepackage{appendix}
\usepackage{subfigure}
\usepackage{ulem}

\begin{document}
\title{Hidden Comet-Tails of Marine Snow Impede Ocean-based Carbon Sequestration}

\author{Rahul Chajwa $^{1}$, Eliott Flaum$^{1}$, Kay D. Bidle $^{5}$, Benjamin Van Mooy $^{6}$,  Manu Prakash $^{1,2,3,4, *}$. }
\affiliation{$^1$Department of Bioengineering, $^2$ Department of Biology, $^3$Department of Ocean, $^4$Woods Institute of the Environment, Stanford University, Stanford CA 94305 USA }
\affiliation{$^{5}$Department of Marine and Coastal Science, Rutgers University, New Brunswick, NJ, 08901, USA}
\affiliation{$^{6}$Woods Hole Oceanographic Institution, Woods Hole, MA, USA}
\affiliation{* To whom correspondence should be addressed: manup@stanford.edu}
\begin{abstract}

Global carbon-cycle on our planet ties together the living and the non-living world, coupling ecosystem function to our climate. Gravity driven downward flux of carbon in our oceans in the form of \textit{marine snow}, commonly referred to as biological pump directly regulates our climate. Multi-scale nature of this phenomena, biological complexity of the marine snow particles and lack of direct observations of sedimentation fundamentally limits a mechanistic understanding of this downward flux. The absence of a physics based understanding of sedimentation of these multi-phase particles in a spatially and temporally heterogeneous ocean adds significant uncertainty in our carbon flux predictions. Using a newly invented scale-free vertical tracking microscopy, we measure for the first time, the microscopic sedimentation and detailed fluid-structure dynamics of marine snow aggregates in field settings. The microscopically resolved in-situ PIV of large number of field-collected marine snow reveals a comet tail like flow morphology that is universal across a range of hydrodynamic fingerprints. Based on this dataset, we construct a reduced order model of Stokesian sedimentation and viscoelastic distortions of mucus to understand the sinking speeds and tail lengths of marine snow dressed in mucus. We find that the presence of these mucus-tails doubles the mean residence time of marine snow in the upper ocean, reducing overall carbon sequestration due to microbial remineralization. We set forth a theoretical framework within which to understand marine snow sinking flux, paving the way towards a predictive understanding of this crucial transport phenomena in the open ocean.



\end{abstract}
\maketitle

\section*{Introduction}
Oceans are the most dominant open reservoir and sink of carbon on our planet \cite{DeVries2022} having absorbed roughly 30\% of the anthropogenically released carbon since industrialization \cite{Wanninkhof2019}. 
A fraction of the absorbed carbon sequesters to the bottom of the ocean \cite{MARTIN1987267}. Of the various known pathways of this vertical transport phenomena, sedimentation of organic carbon in the form of marine snow is the most consequential \cite{ALLDREDGE198841, Boyd2019, Omand2020, Rcarson, Silver_2015, Ben2015}. 
The multiphase nature of marine snow, arising from biological and physical complexity of these particles, can be attributed to multiple length and time scales that plays in this phenomena [Figure 1 A \& B] \cite{ALLDREDGE198841, Burd2009}.  
Although vital to our understanding of current and future climate, the lack of predictive understanding of marine snow sedimentation \cite{SANDERS2014200} manifests itself as significant uncertainty in climate model \cite{Kriest2008, Henson2022}.



Phytoplankton in the upper sunlit layer of the ocean initiates energy flow in marine food-webs \cite{STOCK20141} by converting dissolved CO$_{2}$ into organic carbon \cite{Boyd2019, Falkowski1998, DELAROCHA201493}. With gravity perpetually acting on marine ecosystems \cite{Larson2022.08.19.504465}, a fraction of this organic matter and energy is pumped from the sunlit ecosystem into the abyss \cite{MARTIN1987267, MARTIN2011338}, where it helps sustain benthic life \cite{ALLDREDGE198841,murray1899}. This sequestered carbon decouples from the atmosphere for time-scales ranging from millennium to geologic \cite{Sigman2000, Ducklow2001}, thanks to the unison of gravitational sinking \cite{Eppley1979, SR2001, Witten2017} and thermohaline circulation \cite{STOCKER2000301, Ducklow2001, Wallace1997}. This biologically mediated flux of carbon in the oceans, called the biological pump \cite{Ducklow2001,Stukel2023, DELAROCHA201493}, is conjectured as a key feedback mechanism in both the long-time scale glacial cycles \cite{Sigman2000} and short time-scale variations in global temperature \cite{Cheng_2022, Martin1994}. 
\begin{figure*}[t]
      \begin{center}
      \includegraphics[width=17.5 cm]{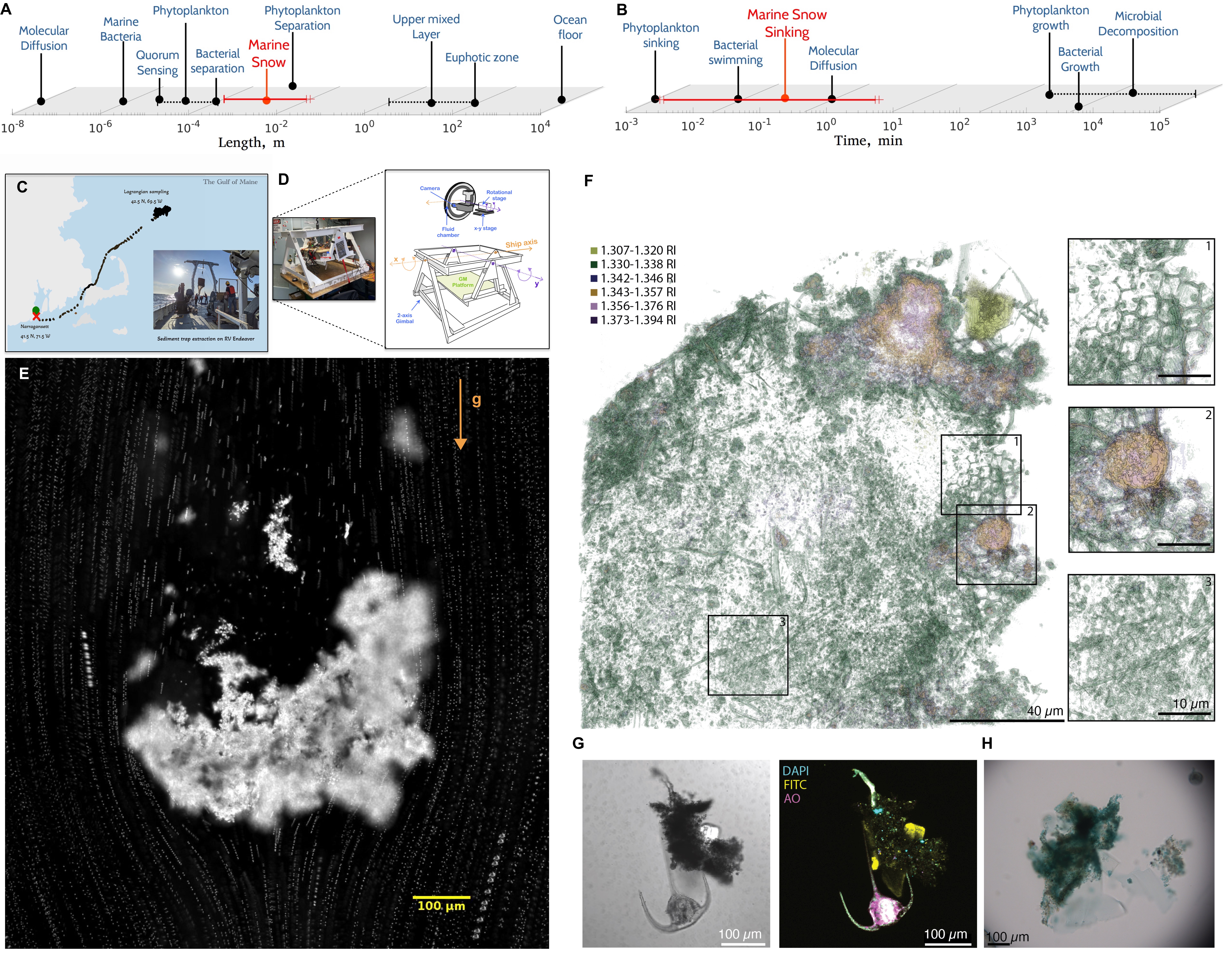}
      \caption{\label{Fig1} \textbf{Marine snow sedimentation in field setting:} A) \& B) The depiction of the multiple length and time scales spanning the biological pump based on previous studies [see Supplementary]. Marine snow exhibits a broad range of sizes and sedimentation time scale. The time scale of sinking is defined as the time it take for marine snow to sink by its own body length based on \cite{LAMPITT2001160}. C) Expedition route to the Gulf of Maine and Lagrangian sampling on the map; inset shows a picture of the sediment trap sampling on research vessel R/V Endeavor. D) Experimental setup of Gravity Machine on a mechanical 2-axis Gimbal hanging on RV-Endeavor in the open ocean. E) Flow-trace image of  sinking marine snow. F) Holotomography imaging using the Tomocube Microscope of a marine snow aggregate collected at 150m on RIPPLE shows the variance of refractive indices within a single particle. G) The inhomogeneous biotic nature of marine snow is depicted in both brightfield and confocal imaging of a marine snow particle, where an embedded dinoflagellate can be seen. The DAPI, FITC and AO channels all emit signal visible from auto-fluorescence of the particle. H) Alcian blue staining of marine snow from an 80m sediment trap to visualize Transparent Exopolymer Particles (TEP) as a representation of polysaccharides present in the marine snow particles. The alcian blue stain is visible throughout the particle.
      }
     \end{center}
\end{figure*}



Single phytoplankton cells, even if they continuously sink, would take about a year to reach the bottom of the ocean; observations however, suggest much smaller time-scale \cite{LAMPITT2001160}. Physical agglomeration \cite{Burd2009} can dramatically increase sinking speed by an order of magnitude, due to increased size \cite{Stokes1851} and reduced drag \cite{chajwa2020}. In addition, zooplankton can enhance particulate density by compactification to form fecal pellets. These factors introduce uncertainty in the sinking speed of marine snow \cite{IVERSEN2020102445}, giving rise to a distribution of size, shape and density. Quantitative measurements of sedimentation traps across various depths in the ocean highlighted the dramatic reduction in carbon flux as a function of depth in our oceans \cite{MARTIN1987267}. This empirical relationship between organic matter flux as a function of depth -- Martin's curve \cite{Martin_1987, Lauderdale2021} is  widely utilized in our current climate models. The microscopic origin of this empirical relationship between organic matter flux as a function of depth remains largely unknown and its universality remains speculative \cite{Lauderdale2021}. 


Due to the paramount nature of these field collected particles, previous attempts have been made to quantify the biochemical \cite{Bidle2015} and physical nature \cite{Ben2015} of these particles. 
Conventional field observations using sediment traps \cite{MARTIN1987267} gives bulk estimates of the exported carbon, while in-situ underwater measurements provide useful size and shape statistics \cite{Alldredge1988, Picheral2010, Trudnowska2021} a one-to-one structure-sedimentation map of marine snow remains elusive \cite{Giering2022}. Since the processes central to the formation, sinking and remineralization of marine snow occur at microscopic scales in the upper layer of the ocean, wider calls have been made previously to highlight this gap in our understanding of marine snow \cite{LAMPITT2001160, SANDERS2014200}.


In this article, we approach the central issue of ocean carbon sequestration by fusing the classic sediment trap sampling \cite{MARTIN1987267} with the state of the art scale-free tracking microscopy called Gravity Machine \cite{Krishnamurthy_2020} to experimentally explore the sedimentation  dynamics and the flow morphology of marine snow aggregates in a field setting, via an expedition in the Gulf of Maine aboard the R/V Endeavor as part of the Research at the Interface of Phytoplankton, Particles and Lipid Export (RIPPLE) expedition. The sedimentation statistics measured using gravity machine offers multi-scale data-sets of marine snow observed at microscopic resolution falling over meter scale. By collecting individual marine snow particles at a depth of 80m and immediately observing sedimentation dynamics of these particles - we acquire the largest database of flow microscopy of sedimenting marine snow at the sea and discover hidden viscoelastic degrees of freedom in the form of mucus, that significantly modifies the fluid-structure interactions and sedimentation velocity of marine snow. Our micro-scale observations and theoretical framework allows sedimentation based visco-elastic rheology of marine snow, measured at sea, for the first time. We find that the presence of mucus comet tails crucially slows down marine snow, bringing some particles to a stand still. This mucus based impedence nearly doubles the residence time of marine snow aggregates in the upper productive layer of the ocean, potentially facilitating rapid remineralization by microbes and zooplankton. We discover that the multiphase nature of comet-like mucus tails in mairne snow has a consequential impact on  ocean-based carbon sequestration.

\begin{figure*}[t]
      \centering
      \includegraphics[width=16cm]{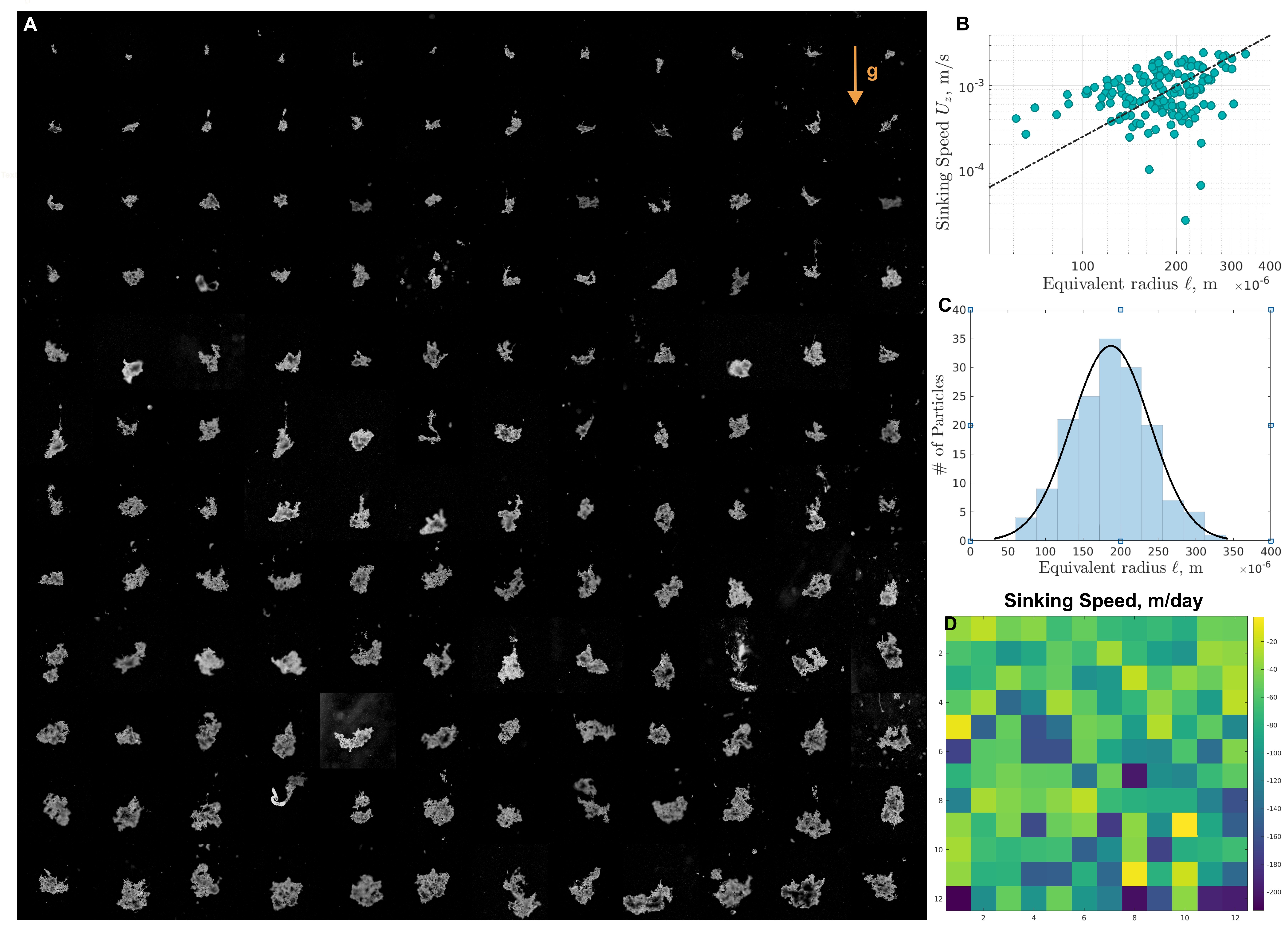}
      \caption{\label{Fig3} \textbf{The zoo of sinking aggregates:}
      D) The zoo of aggregates in their stable sinking configuration. The aggregate in Fig 1.E is (row , column) = (9 , 5). Each image panel is 1.27 x 1.27 mm$^2$. B) log-log plot of sinking speed as a function of equivalent radius, with the dotted line representing the Stokes' law fit ($U_{z}\sim l^2$) with a constant effective density $\rho_{sw} + 11.4 Kg/m^{3}$, depicting lack of deterministic Stokes' trend. C) Experimental sample distribution of visible particulate aggregate sizes with normal distribution fit gives mean $\mu = 187.35 \pm 8$ micron and standard deviation $\sigma = 51.93 \pm 6$ micron in the observed sizes. D) Sinking speeds corresponding to D), where each (row , column) in D) corresponds to the same (row , column) in A).
      }
      \label{equation:1}
\end{figure*}

\begin{figure*}[t]
      \begin{center}
      \includegraphics[width=17.5 cm]{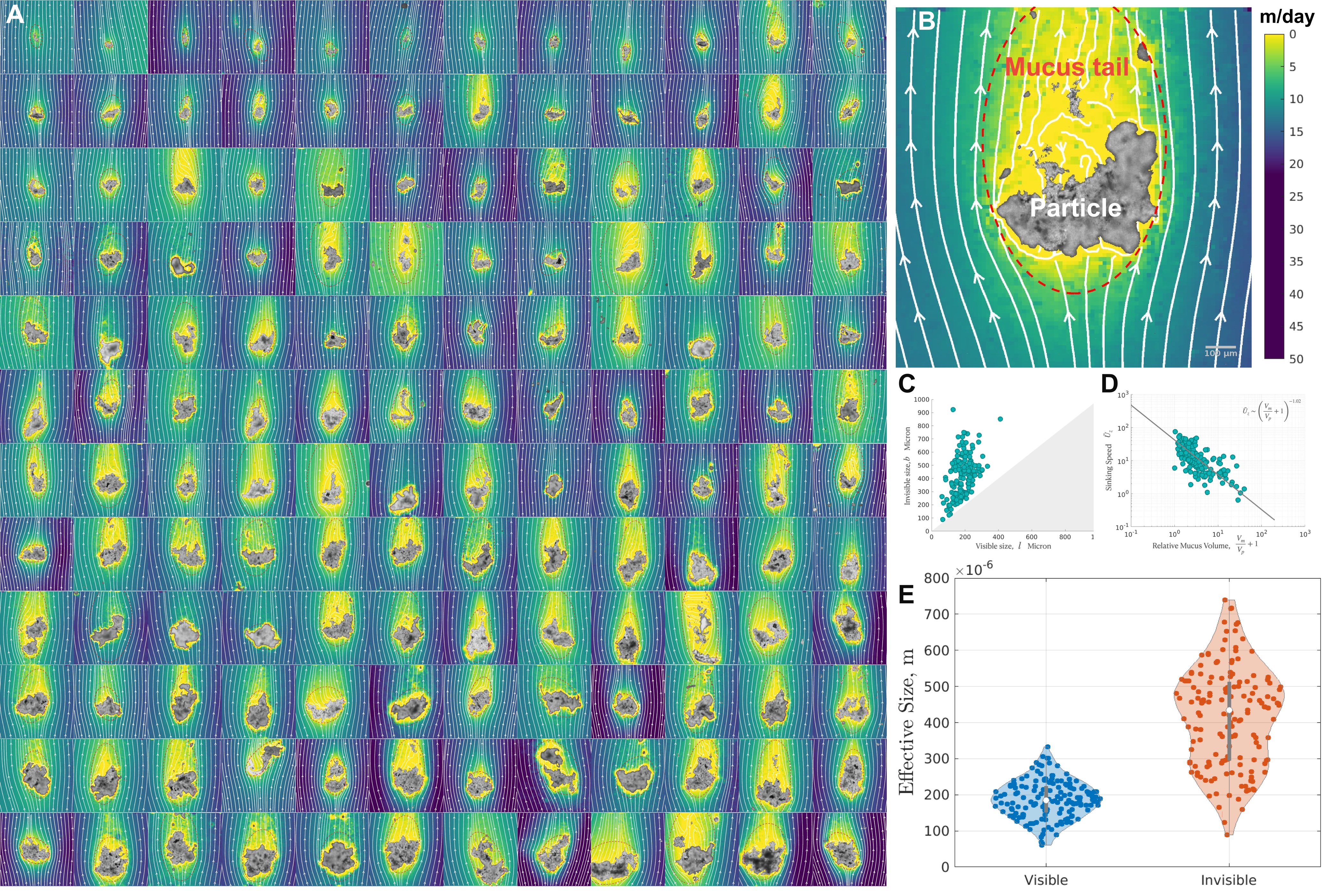} 
      \caption{\label{Fig4} \textbf{Hydrodynamic signature of marine snow:} A) PIV images showing the hydrodynamic signature of the marine snow, making visible the mucus comet-tails. B) Mucus tail around the visible aggregate (row , column) = (9 , 5), same aggregate as Fig.1 E. C) The visible component of marine snow plotted against the invisible mucus degree of freedom. D) Non-dimensional sinking speed \eqref{sinking_ND} plotted against the relative mucus volume $V_{m}/V_{p}$, shows a systematic decrease in sinking speeds as a function of mucus volume with an exponent $\simeq -1$. E) The violin plot of the invisible size of marine snow (red) compared with the distribution of the visible size spectrum (blue).  Accounting for the mucus gives the mean size $\mu_{m} = 415.1 \pm 21 \, \mu$m and standard deviation $\sigma_{m} = 136.3 \pm 15 \, \mu$m [see Fig. 2 C for comparison]
      }
     \end{center}
\end{figure*}
\section*{Results}
\subsection*{Field Observations}

In order to collect fresh marine snow aggregates, an expedition in the Gulf of Maine was planned. The marine snow aggregates were collected in a freely hanging sediment traps (2m diameter) \cite{Ben2015} in the Gulf of Maine ($42.5^{\circ}$ N $69.5^{\circ}$ W), on the research vessel RV Endeavor [see Figure 1C]. In multiple deployments of these traps at depths of 80m for a duration of 24 hours [see supplementary], we acquired marine snow as a sediment. 


Fresh marine snow aggregates were collected on board R/V Endeavor during the RIPPLE expedition in the Gulf of Maine (42.5$^{\circ}$N, 69.5$^{\circ}$W) using freely hanging sediment traps (2m diameter) in multiple deployments at depths of 80m for a duration of 24 hours [see Supplementary]. Particles accumulated in a sediment traps were brought to the surface. The samples were collected and divided using a quantitative splits [see supplementary], and the marine snow particle were kept at the \textit{in-situ} temperature to maintain biotic component of the marine snow alive and active. The particles were loaded immediately in scale-free tracking microscopy wheels and imaged instantly. For a high-density flux of particles, some marine snow particles can accumulate and thus loose their individual identity in sediment trap sampling, and become part of the bulk sediment. We gently re-suspended those particles and focused on the response of the organic debris to the hydrodynamic stresses generated by gravitational sinking \cite{SR2001}. We also ensured a dilute limit in which inter-particle hydrodynamic interactions could be neglected. The ship frame of reference contributes rocking and rolling disturbances to the setup which we minimize by mounting the Gravity Machine on a two axis gimbal [Figure 1 D]. 



Micro-scale sedimentation dynamics of falling marine snow particles was measured in the Gravity Machine \cite{Krishnamurthy_2020}, as soon as the particles are brought to the research vessel [Fig. 1 E]. We directly measured the settling velocity by tracking individual particle in the gravity machine while simultaneously performing high resolution imaging of marine snow and the micro-hydrodynamic flow around it. Plastic bead particles ranging from 700nm to 2 microns are addeded to the fluid for high-resolution particle image velocimetry (PIV). We developed an analysis pipeline that automatically measures the size and velocity statistics of the particles and calculate and analyse the flow fields, allowing us to get a one-to one map between various measurable quantities -- structure, flow morphology and sedimentation velocity. 

\subsection*{Heterogeneous micro-structure of marine snow }

Limited volumetric imaging has previously been performed on marine snow particles in the past [cite].  To better understand the three-dimensional structure of a marine snow particles, we perform the first quantitative phase imaging (Holotomograoghy) in 3D and obtain isolate materials based on refractive index (Fig. 3A). This method allowed us to observe the 3D density and heterogeneity of samples without fixation or stains, thus preserving the cell density and porosity of the aggregates \cite{Park2018}. The heterogeneous structure was clearly depicted, highlighting segments of diatom frustules embedded in the marine snow particle as ballast (Fig. 3A, Supplementary Video 3). Secondly we imaged particles using brightfield and confocal microscopy (Fig. 3B) to observe the presence of biological materials within the sample. Exciting the particle with DAPI, FITC, and AO we observe emission of autofluorescence detecting live cells and associated chloroplasts. Lastly, we imaged with a color camera and brightfield imaging after staining particles with Alcian blue to visualize Transparent Exopolymer Particles (TEP) as a representation of polysaccharides present in the marine snow particles (Fig. 3C). This is commonly used for marine snow particles to reveal polysaccharides which might be surrounding the particle but not visible. In combination, these three imaging datasets created a more complete picture of the composition of marine snow particles. We are able to observe the heterogeneity including the presence of biotic material, ranging from whole cells (Fig. 3B) to fractions of whole cells, both rigid in diatom frustrules (Fig. 1D) and soft in mucus (Fig. 1EF). 

\subsection*{Direct measurement of sedimentation dynamics}
 A predictive understanding of marine snow sedimentation necessitates in-situ measurements of a one-to-one map between the structure of marine snow and the sinking velocity \cite{SANDERS2014200}. 
Thus, we embarked on measuring detailed dynamics of all collected particles while on-board the ship. We track individual marine snow particles in the GM, measuring their sinking speed, and carrying out detailed microscopy at the resolution of 0.828 $\mu$m per pixel at the rate of 5 frames per second. This presents a high resolution zoo of aggregate shapes and sizes [Fig.2 A ], where gravitational torques due to shape polarity has already aligned the principle axis of the aggregate along gravity \cite{Witten_2020}, resulting in a nearly stable orientation over the measurement window of $\simeq 5$ min. In addition, the hydrodynamic levitation directly measures the sinking speed associated with each particle. Plotting sinking speeds as a function of equivalent spherical radius calculated by thresh-holding method \cite{10.3389/fmars.2020.00564} [see Supplementary], shows significant spread, presenting the lack of a deterministic trend [Fig2. B]. We studied a visible spectrum of particle `radius' ranging from $50 - 350 \,\mu$m, with mean radius $187.35 \pm 8 \, \mu$m [Fig.2 C], and present a one-to-one map between detailed structure of marine snow [Fig.1 A] and sedimentation velocity [Fig.2 D], with velocities ranging between $5 - 200$ m/day. In our observations, we occasionally encountered unsteady singular events like merging, slow depletion due to hydrodynamic stresses, and grazing by ciliates and copepod; all of which adds further richness to marine snow sedimentation and likely makes it a highly dynamic process at long-time scales.

\textit{In-situ} observations of marine snow (Fig.1) \cite{McDonnell_2012, Alldredge1988, Picheral2010, Trudnowska2021} along with Martin's curve \cite{Martin_1987}, has allowed meaningful estimates of carbon flux in the ocean. These calculations, however, rely on a deterministic relationship between particle size and its sinking speed \cite{Giering2022}, which has thus far been elusive [Fig. 2 B], presenting a long-standing puzzle in the sedimentation of marine snow \cite{IVERSEN2020102445}. This suggests the presence of unaccounted degrees of freedom in the system, likely coming from incorrect size estimates \cite{Giering2022} and the density fluctuations across particles \cite{MCCAVE1975491}. We seek further cues of any hidden degrees of freedom in the flow field around individual aggregates. 



%





\subsection*{Invisible Comets of Mucus}



The ship environment presents a challenging set-up for extracting clean flow-field from consecutive particle laden images, because of the high-frequency jitters in the equipment, in addition to the canonical rock and roll dynamics. A 2-axis gimbal helped reduce the low frequency structured noise, and the high frequency random noise was removed in post-processing pipeline [see Supplementary]. This allowed for the first micro-scale PIV (Fig. 3 A) and corresponding sedimentation velocity (Fig. 2 D) of marine snow in a field setting, to the best of our knowledge. The measured flow around these sinking aggregates conspicuously displayed the transparent mucus around them. Note that the aggregate in Fig. 3 B is the same as Fig.1 E, with its viscoelastic degrees of freedom visible. Alcian blue staining (Fig. 1 H) confirms that this mucus halo consisted of Transparent Exo-Polymers, which is known to have a biological origin in algal blooms and plays a key role in particle agglomeration \cite{PASSOW2002287, Burd2009}. This presents a multi-phase fluid-structure interaction problem.

The detailed PIV enabled visualization of the comet like morphology of marine snow which is universal across all particle shape and sizes. We found that the amount of mucus that an aggregate is dressed with is independent of the size of the visible aggregate (Fig. 3 C). This invisible degree of freedom significantly changes the effective size of the aggregate, sets a different fluid-mechanical boundary condition, and reduces the effective density. This is evident in the inverse relationship between the relative volume of mucus to particle and the settling velocity (Fig. 3 D). The invisible size spectrum is dramatically different from the visible spectrum, with a mean $415 \pm 21 \, \mu$m,  which is more than twice the mean size of the visible part of marine snow (Fig. 2 C). 



\begin{figure*}[t]
      \begin{center}
      \includegraphics[width=17.5 cm]{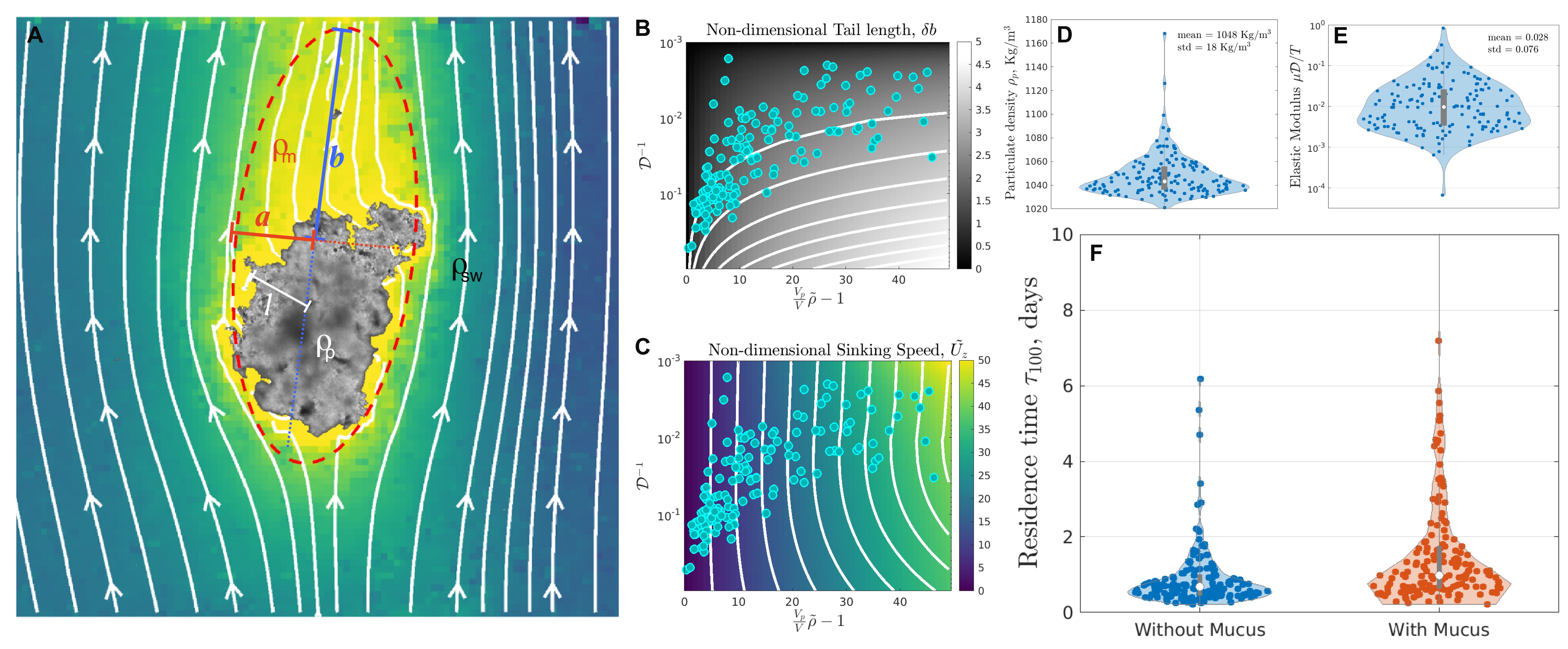}
      \caption{\label{Fig5} \textbf{Two-phase model and sedimentation based rheology of marine snow:} A) Measured flow field around a sinking particle corresponding to (8,3) in Fig.2D. An approximately ellipsoidal mucus halo enclosing the visible particulate aggregate, defines the various densities $\rho_{m}$, $\rho_{p}$, $\rho_{sw}$ and length scales $a$, $b$, $l$ in the mucus comet-tail. B) The steady state solution of equation \eqref{KV_nondimensional} gives the mucus tail length $\delta b$ as a function of the internal structural degrees of freedom of marine snow (${V_{p}}/{V}$, $\tilde{\rho}$, $\mathcal{D}$). White curves are the contours of constant tail length. C) Steady state solution of equation \eqref{sinking_ND} gives the vertical sinking speed as a function of (${V_{p}}/{V}$, $\tilde{\rho}$, $\mathcal{D}$). White curves are constant speed contours. D) The corrected distribution for particulate density after accounting for the mucus. E) The distribution of elastic modulus of the mucus by measuring the distorsion and sinking speed, presenting a sedimentation based rheology. F) The residence time distribution of marine snow in upper 100m in the ocean, for case when there is mucus (red) and without mucus (blue) [see Supplementary for details]}
     \end{center}
\end{figure*}

\subsection*{Two phase model of marine snow \textit{dressed} in mucus}

Based on our field observations and analysis of freshly collected marine snow aggregates in the Gulf of Maine, we constructed a minimal model for the Stokesian sinking of marine snow with mucus halo by analytically treating the halos as an deformable ellipsoid \cite{McNown1950} with semi-major and semi-minor axes $b$ and $a$, respectively, and density $\rho_{m}$. The visible particulate aggregate with effective radius $l$ has a density $\rho_p$ greater than $\rho_{m}$ [Fig. 4 A]. Here we treat marine snow as a two-phase aggregate with visible particulate aggregate embedded in the invisible mucus halo, sedimenting under its own weight. Hydrodynamic self-stress generated due to sedimentation would perturb the shape of mucus - further deforming and hence influencing the steady state sinking velocity of these particles. We construct the minimal model that accounts for this feedback between deformation and sedimentation and elucidate the microscopic origins of marine snow sedimentation dynamics.



We always found that the particulate matter is present at the bottom of the mucus halo, thus marine snow with mucus halo is effectively a bottom heavy sedimenting particle \cite{Witten2017, nissanka_ma_burton_2023}. 
In the ambit of linear response, the shape degree of freedom $\delta \vect{b} \equiv (b/b_{0} - 1)  \hat{z}$ is assumed to evolve viscoelastically through the Kelvin-Voigt (KV) model \cite{Christensen1984} and the translational degree of freedom $\vect{X}$ is governed by the Stokesian mobility relation \cite{Happel1983, KIM1991} with a shape dependent mobility $\barbelow{\barbelow{\mathbb{M}}}(\vect{\delta b})$  \cite{McNown1950, Witten_2020}, giving rise to the coupled equations


\begin{equation}
     \dot{\barbelow{\vect{\delta b}}} + \frac{E}{\mu'} \barbelow{\vect{\delta b}} = \frac{\alpha}{\mu'} \barbelow{\barbelow{\vect{\sigma}}} \cdot \barbelow{\vect{b}}
     \label{KV}
\end{equation}

\begin{equation}
    \barbelow{\dot{\vect{X}}} = \barbelow{\barbelow{\mathbb{M}}}(\vect{\delta b}) \cdot \barbelow{\vect{F}}
    \label{mobility}
\end{equation}


 for shape and sinking dynamics respectively. In \eqref{KV} $\vect{\sigma} = (\vect{\nabla v} + \vect{\nabla v} ^{T})/2$ is the symmetric part of the velocity gradient generated by a force monopole of strength $\vect{F}$, which is assumed to be located at the origin of visible particulate aggregate, due to its relatively dominant contribution to weight as compared to mucus \cite{Guo2021}. $\vect{F}$ induces translational dynamics of the center of gravity $\vect{X}$ of marine snow through the mobility relation \eqref{mobility}. In equation $\eqref{KV}$ the second term on the left captures the visco-elastic relaxation, and the term on the right presents a one-way hydrodynamic coupling between the KV dimer located at the origin of mucus halo and sedimentation induced flow $\vect{v}$, to leading orders in velocity gradient.  
 The phenomenological parameter $\alpha$ can, in general, depend on the shape of the visco-elastic halo, and determining it requires solving the detailed fluid mechanical problem \cite{doi:10.1146/annurev-fluid-010719-060107}. For the sake of simplicity in the present analysis, we fix $\alpha = \mu$ on dimensional grounds. When the extensional axis is aligned with the gravity axis, thanks to the polarity of the marine snow, the Stokesian hydrodynamic stress induced by $\vect{F}$ along $\vect{b}$, to leading order in $l/a_{0}$ becomes [see Supplementary text]
\begin{equation}
    \barbelow{\barbelow{\vect{\sigma}}} \cdot \barbelow{\vect{b}} = \frac{2}{3} \frac{a_{0} (\rho_{sw} - \rho_{m}) \vect{g}}{(1 + \delta b)^2} \left( \frac{V_p}{V} \tilde{\rho} - 1\right)
    \label{reduced_selfstress}
\end{equation}
 We have defined $\tilde{\rho} \equiv (\rho_p - \rho_m)/(\rho_{sw} - \rho_m)$, which lumps the sea water density $\rho_{sw}$, mucus density $\rho_{m}$ and aggregate density $\rho_{p}$ into a single non-dimensional density parameter; with $V_{p}$ and $V$ being the volume of the particulate aggregate and the net volume of the marine snow respectively. We consider a volume conserving mucus and an isotropic initial condition with non-zero mucus volume, $ b_0 = a_{0} > l$. Combining \eqref{KV} and \eqref{reduced_selfstress} and non-dimensionalizing \eqref{KV} and \eqref{mobility} using length scale $L= a_{0}$ and sedimentation time scale $T = 9 \mu/ 2 g (\rho_{sw} - \rho_{m}) a_{0}$ for $ \rho_{m} \neq \rho_{sw}$ gives the dynamical equations for longitudinal deformation $\delta b$ and vertical depth $z$ scaled by $a_{0}$,

\begin{equation}
    \dot{\delta b } + \frac{\mu}{\mu'} \, \mathcal{D} \, \delta b = 3 \frac{\mu}{\mu'}  \left(\frac{V_{p}}{V} \tilde{\rho} - 1 \right) \frac{1}{(1 + \delta b)^2},
    \label{KV_nondimensional}
\end{equation}
\begin{equation}
    \dot{z} \, =  \left( \frac{V_{p}}{V} \tilde{\rho} - 1 \right) \frac{\mathbb{X}^{-1}(e)}{ 1 + \delta b}.
    \label{sinking_ND}
\end{equation}


In \eqref{KV} $\mathcal{D} \equiv 9E/[2g(\rho_{sw}-\rho_{m})a_{0} ]$ is the ratio of elastic to hydrodynamic stress response, the magnitude of which is set by the biology of mucus secretion in phytoplankton \cite{PASSOW2002287, Meng2016, Bidle2015}. The scalar mobility of the spheroid with its symmetry axis aligned with gravity, $\mathbb{X}^{-1}(e) \equiv {3}\left[ (1 + e^2)\ln \left( {1 + e}/{1 - e}\right) - 2 e\right ]/{8 e^3}$ captures shape dependent sinking as a function of eccentricity $e = \sqrt{ 1 - (1+\delta b)^{-3} }$. Equation \eqref{sinking_ND} gives the condition for non-zero sinking of the marine snow $\tilde{\rho} \geq V/V_{p}$
implying that marine snow with larger mucus halos falls slower, potentially reducing the net carbon flux in the ocean \cite{PASSOW2002287}. 



We solve \eqref{KV_nondimensional} and \eqref{sinking_ND} in the steady state approximation of the mucus comet tail $\dot{\delta b} = 0$. A stable fixed point is guaranteed in \eqref{KV_nondimensional} and the steady state tail length has a closed analytical expression [see Supplementary text] which is plotted as a function of $\mathcal{D}$, $V_{p}/V$ and $\tilde{\rho}$ [Fig. 3 B]. Asymptotic analysis of the steady state yields two scaling regimes for very small and very large deformations
\begin{equation}
  \delta b \sim \begin{cases}
               \left(\frac{V_{p}}{V} \tilde{\rho} - 1\right) \mathcal{D}^{-1} \,\, , \,\,\delta b \ll 1\\
              \left(\frac{V_{p}}{V}\tilde{\rho} - 1\right)^{1/3} \mathcal{D}^{-1/3} \, \, , \,\,\delta b \gg 1 
            \end{cases}
            \label{extension_scaling}
\end{equation}
where the viscosity sets the time scale to achieve the steady state; however, \eqref{extension_scaling} is independent of $\mu$ and $\mu'$. Figure 4 C shows the steady state sinking velocity \eqref{sinking_ND} as a function of $\mathcal{D}$, $V_{p}/V$ and $\tilde{\rho}$. Comparison between theory and field data allows us to get the density distribution of the visible particulate matter [Fig. 4 D], which further enables the sedimentation based measurement of the Elastic modulus of mucus [see Supplementary] with the mean value of $0.028$ Pascal [Fig. 4 E]. Viscoelastic rheology of xanthan gum in salt solutions provides useful insights into hydrodynamic nature of marine mucus, while in-situ measurements remain limited \cite{Mrokowska2022, Jenkinson1986}. Here we present a sedimentation based measurement of the storage modulus of fresh marine snow mucus in field setting \cite{PASSOW2002287}.

Our field experiments and theoretical framework shows that sinking speed of two-phase particle is inversely proportional to the relative volume of mucus in marine snow [Fig. 3 D], in accord with equation \eqref{sinking_ND}. This mucus induced impedance in marine snow reduces the sedimentation flux and increases the residence time of mucus rich aggregates in our data-set. The estimate of residence time shows that presence of mucus almost doubled the mean residence time of marine snow in the upper 100m of the ocean  (1.92 days)  as compared to that without mucus (0.9 day)  [see Fig. 4 F and Supplementary], assuming a steady sinking state.

\section*{Summary and Future Challenges}

The advances in our understanding of ecosystems in open oceans, and of marine snow in particular, has historically developed in tandem with the advances in marine technology. Our study makes use of a recent technological development in scale-free vertical tracking microscopy \cite{Krishnamurthy_2020} implemented on a research vessel allows us to shed light on the essential dynamical aspects of marine snow sedimentation. To the best of our knowledge, we present the first microscopic window into the sedimentation physics of these complex multi-phase particles and highlight significant consequences to carbon flux in our oceans. Linking this transport phenomena that is governed by micro-scale parameters to larger scale ocean carbon flux dynamics \cite{Martin_1987} is crucial for understanding tipping points and balancing the carbon budget in current ocean based carbon sequestration estimates \cite{Boyd2019}. 

Our work presents the largest known dataset of micro-scale hydrodynamics of marine snow sedimentation. By directly measuring the sinking velocities and detailed flows around individual marine snow particles at sea moments after they are brought on board, we discovered long comet-like mucus tails arising from TEP around marine snow, whose accurate form and eventual role in sedimentation is otherwise invisible in traditional imaging. These comet tails endow marine snow with a modified sinking speed, dramatically different from estimates made just by looking at the visible size of the particle. Our observations provide a new theoretical framework, based on Stokesian dynamics, in which we include this previously invisible degree of freedom and we present a modified stokes law for these compound particles. We present a combination of field experiments and theory that also allows us to carry out sedimentation based rheology of marine snow in the field, enabling estimates of in-situ visco-elastic properties of marine snow. 

 It has been well known that marine snow particles contain an invisible (transparent) mucus layer that impacts the flocculation of diatom and cocolithopore blooms \cite{PASSOW2002287,Laber2018}. The carbon content of TEP is comparable to that of the particular matter \cite{Mari1999}.
The TEP production is shown to increase linearly with CO$_{2}$ uptake \cite{Engel2002}, and in response to both abiotic and biotic stressor in phytoplankton \cite{Bidle2015}; climate change can directly tune carbon sequestration with the TEP knob \cite{Arrigo2007}. However, the surprising role of this mucus in altering the sedimentation dynamics of marine snow has remained unknown. 

In our net trap samples with over hundred particles imaged directly via this technique, we find that presence of mucus almost doubled the residence time of marine snow in the euphotic zone, nearly halting some particles to a stand still. We demonstrate that this slow-down occurs due to mucus comet tails significantly changing the hydrodynamic footprint of a sinking particle. Our work emphasizes the critical role mucus comet tails can play in reducing the sedimentation velocity of marine snow particles. Taken in light of the fact that many of the current estimates of marine snow sedimentation rely on ``visible" size distribution and do not account for much of the complexity of sedimentation of these particles - our findings can help reconcile significant discrepancies in flux estimates reported previously. We corroborate our findings with 3D volumetric imaging of marine snow particles and detect various ballast including diatom and foram shells - that further highlight the heterogeneous nature of marine snow particles and hence variation in it's density at large. With a unique microscopic observational window - we provide the first multi-scale dataset with imaging resolution of less than a micron on a sedimentation dynamics that occur on meter scale.
Given the many-particle character of agglomeration of organic matter to form marine snow \cite{Burd2009}, we expect that a probability distribution in $(\mathcal{D}, V_{p}/V, \tilde{\rho})$ parameter space [Fig. 4 C] governs the net flux of matter through a horizontal plane in the ocean. A detailed knowledge of this distribution, which is necessary for estimating carbon flux in the ocean, is currently missing. Future expeditions will hopefully fill this gap in available data. We believe this approach will pave the way to linking micro-scale in-situ observations of marine snow to understanding the macro-scale biological pump from the bottom-up, and bringing marine snow phenomena in the ambit of soft matter physics. 

The fact that these mucus heavy aggregates host microbes and plankton (both swimming and non-swimming) endows marine snow with an active component. Although imaging tools presented here allow us to investigate these processes - they fall outside the scope of current work and the biological dynamics of these processes and it's role in sedimentation will be investigated in future expeditions. Although many of the marine snow particles appear to be porous - we find that most often mucus clogs these pores, modifying bacterial motility and degradation\cite{10.1093/pnasnexus/pgac311}.




The rapidly changing climate \cite{Cheng_2022}, necessitates improved observations and predictive understanding of oceanic carbon flux. Our work currently describes the complex sedimentation dynamics of marine snow particles. Combined with mechanisms and rates of marine snow formation and their remineralization by microbes, this framework has the potential to provide a fundamental bottom-up description for 1-D flux models such as the Martin's curve \cite{Martin_1987}. Currently it is estimated that 30\% of anthropogenic carbon is sequestered by the ocean via biological pump. We observe a 100 \% increase in residence time of mucus heavy marine snow particles, likely facilitating remineralization by microbes and sinificantly altering the overall flux that can be sequestered by this mechanism. Given the current uncertainty in biological pump in a changing global climate \cite{Lauderdale2021}, it is imperative to underpin the microbiology and micro-physics of marine snow. Direct in-situ and on-vessel field setting measurements of broad range of marine snow particles provides a promising approach to predictively understanding the biological pump.


\medskip
\begin{acknowledgements}
    We thank the captain and crew of the R/V Endeavor, as well as the entire RIPPLE science team, including Helen F Fredricks for her efforts and expertise in marine snow particle splits, and thank the gravity machine team, especially Hongquan Li and Melanie Hannabelle for their technical support. RC acknowledges support from the Human Frontier Science Program Organisation, the Stanford Bio-X travel grant, and the discussions in the Active matter in Complex Environments conference at the Aspen Center for Physics  Jan 2022. This work was enabled by a NSF Growing Convergence Research (GCR) grant to KDB, BVM and MP. We thank the entire GCR research team for their collaborative, interactive and inspiring discussions.
    MP further acknowledges financial support from Schmidt Futures Innovation Fellowship, Moore Foundation and CZI BioHub investigator funding. 
\end{acknowledgements}



\bibliography{marine_snow}

\begin{thebibliography}{10}

\bibitem{DeVries2022}
T.~DeVries, ``The ocean carbon cycle,'' {\em Annual Review of Environment and
  Resources}, vol.~47, no.~1, pp.~317--341, 2022.

\bibitem{Wanninkhof2019}
N.~G. et~al., ``The oceanic sink for anthropogenic co2 from 1994 to 2007,''
  {\em Science}, vol.~363, no.~6432, pp.~1193--1199, 2019.

\bibitem{MARTIN1987267}
J.~H. Martin, G.~A. Knauer, D.~M. Karl, and W.~W. Broenkow, ``Vertex: carbon
  cycling in the northeast pacific,'' {\em Deep Sea Research Part A.
  Oceanographic Research Papers}, vol.~34, no.~2, pp.~267--285, 1987.

\bibitem{ALLDREDGE198841}
A.~L. Alldredge and M.~W. Silver, ``Characteristics, dynamics and significance
  of marine snow,'' {\em Progress in Oceanography}, vol.~20, no.~1, pp.~41--82,
  1988.

\bibitem{Boyd2019}
P.~W. Boyd, H.~Claustre, M.~Levy, D.~A. Siegel, and T.~Weber, ``Multi-faceted
  particle pumps drive carbon sequestration in the ocean,'' {\em Nature},
  vol.~568, pp.~327--335, Apr. 2019.

\bibitem{Omand2020}
M.~M. Omand, R.~Govindarajan, J.~He, and A.~Mahadevan, ``Sinking flux of
  particulate organic matter in the oceans: Sensitivity to particle
  characteristics,'' {\em Scientific Reports}, vol.~10, Mar. 2020.

\bibitem{Rcarson}
R.~Carson, {\em The sea around us}.
\newblock New York: Oxford University Press, 2003.

\bibitem{Silver_2015}
M.~Silver, ``Marine snow: A brief historical sketch,'' {\em Limnology and
  Oceanography Bulletin}, vol.~24, pp.~5--10, jan 2015.

\bibitem{Ben2015}
J.~R. Collins, B.~R. Edwards, K.~Thamatrakoln, J.~E. Ossolinski, G.~R.
  DiTullio, K.~D. Bidle, S.~C. Doney, and B.~A.~S. Van~Mooy, ``The multiple
  fates of sinking particles in the north atlantic ocean,'' {\em Global
  Biogeochemical Cycles}, vol.~29, no.~9, pp.~1471--1494, 2015.

\bibitem{Burd2009}
A.~B. Burd and G.~A. Jackson, ``Particle aggregation,'' {\em Annual Review of
  Marine Science}, vol.~1, pp.~65--90, Jan. 2009.

\bibitem{SANDERS2014200}
R.~S. et~al., ``The biological carbon pump in the north atlantic,'' {\em
  Progress in Oceanography}, vol.~129, pp.~200--218, 2014.
\newblock North Atlantic Ecosystems, the role of climate and anthropogenic
  forcing on their structure and function.

\bibitem{Kriest2008}
I.~Kriest and A.~Oschlies, ``On the treatment of particulate organic matter
  sinking in large-scale models of marine biogeochemical cycles,'' {\em
  Biogeosciences}, vol.~5, pp.~55--72, Jan. 2008.

\bibitem{Henson2022}
S.~A. Henson, C.~Laufk\"{o}tter, S.~Leung, S.~L.~C. Giering, H.~I. Palevsky,
  and E.~L. Cavan, ``Uncertain response of ocean biological carbon export in a
  changing world,'' {\em Nature Geoscience}, vol.~15, pp.~248--254, Apr. 2022.

\bibitem{STOCK20141}
C.~A. Stock, J.~P. Dunne, and J.~G. John, ``Global-scale carbon and energy
  flows through the marine planktonic food web: An analysis with a coupled
  physical–biological model,'' {\em Progress in Oceanography}, vol.~120,
  pp.~1--28, 2014.

\bibitem{Falkowski1998}
P.~G. Falkowski, R.~T. Barber, and V.~Smetacek, ``Biogeochemical controls and
  feedbacks on ocean primary production,'' {\em Science}, vol.~281,
  pp.~200--206, July 1998.

\bibitem{DELAROCHA201493}
C.~{De La Rocha} and U.~Passow, ``8.4 - the biological pump,'' in {\em Treatise
  on Geochemistry (Second Edition)} (H.~D. Holland and K.~K. Turekian, eds.),
  pp.~93--122, Oxford: Elsevier, second edition~ed., 2014.

\bibitem{Larson2022.08.19.504465}
A.~G. Larson, R.~Chajwa, H.~Li, and M.~Prakash, ``A topologically complex
  cytoplasm enables inflation-based escape from a gravity trap,'' {\em
  bioRxiv}, 2022.

\bibitem{MARTIN2011338}
P.~M. et~al., ``Export and mesopelagic particle flux during a north atlantic
  spring diatom bloom,'' {\em Deep Sea Research Part I: Oceanographic Research
  Papers}, vol.~58, no.~4, pp.~338--349, 2011.

\bibitem{murray1899}
S.~J. Murray, ``Address to the geographical section of the british association,
  1899,'' {\em Scottish Geographical Magazine}, vol.~15, no.~10, pp.~505--522,
  1899.

\bibitem{Sigman2000}
D.~M. Sigman and E.~A. Boyle, ``Glacial/interglacial variations in atmospheric
  carbon dioxide,'' {\em Nature}, vol.~407, pp.~859--869, Oct. 2000.

\bibitem{Ducklow2001}
H.~Ducklow, D.~Steinberg, and K.~Buesseler, ``Upper ocean carbon export and the
  biological pump,'' {\em Oceanography}, vol.~14, no.~4, pp.~50--58, 2001.

\bibitem{Eppley1979}
R.~W. Eppley and B.~J. Peterson, ``Particulate organic matter flux and
  planktonic new production in the deep ocean,'' {\em Nature}, vol.~282,
  pp.~677--680, Dec. 1979.

\bibitem{SR2001}
S.~Ramaswamy, ``Issues in the statistical mechanics of steady sedimentation,''
  {\em Advances in Physics}, vol.~50, no.~3, pp.~297--341, 2001.

\bibitem{Witten2017}
T.~Goldfriend, H.~Diamant, and T.~A. Witten, ``Screening, hyperuniformity, and
  instability in the sedimentation of irregular objects,'' {\em Phys. Rev.
  Lett.}, vol.~118, p.~158005, Apr 2017.

\bibitem{STOCKER2000301}
T.~F. Stocker, ``Past and future reorganizations in the climate system,'' {\em
  Quaternary Science Reviews}, vol.~19, no.~1, pp.~301--319, 2000.

\bibitem{Wallace1997}
W.~S. Broecker, ``Thermohaline circulation, the achilles heel of our climate
  system: Will man-made co2 upset the current balance?,'' {\em Science},
  vol.~278, no.~5343, pp.~1582--1588, 1997.

\bibitem{Stukel2023}
M.~R. Stukel, J.~P. Irving, T.~B. Kelly, M.~D. Ohman, C.~K. Fender, and
  N.~Yingling, ``Carbon sequestration by multiple biological pump pathways in a
  coastal upwelling biome,'' {\em Nature Communications}, vol.~14, Apr. 2023.

\bibitem{Cheng_2022}
L.~Cheng, K.~von Schuckmann, J.~P. Abraham, K.~E. Trenberth, M.~E. Mann,
  L.~Zanna, M.~H. England, J.~D. Zika, J.~T. Fasullo, Y.~Yu, Y.~Pan, J.~Zhu,
  E.~R. Newsom, B.~Bronselaer, and X.~Lin, ``Past and future ocean warming,''
  {\em Nature Reviews Earth and Environment}, vol.~3, pp.~776--794, oct 2022.

\bibitem{Martin1994}
J.~H.~M. et~al., ``Testing the iron hypothesis in ecosystems of the equatorial
  pacific ocean,'' {\em Nature}, vol.~371, pp.~123--129, Sept. 1994.

\bibitem{LAMPITT2001160}
R.~Lampitt, ``Marine snow,'' in {\em Encyclopedia of Ocean Sciences (Third
  Edition)} (J.~K. Cochran, H.~J. Bokuniewicz, and P.~L. Yager, eds.),
  pp.~160--169, Oxford: Academic Press, third edition~ed., 2001.

\bibitem{Stokes1851}
G.~{Stokes}, ``{On the Effect of the Internal Friction of Fluids on the Motion
  of Pendulums},'' {\em Transactions of the Cambridge Philosophical Society},
  vol.~9, p.~8, Jan. 1851.

\bibitem{chajwa2020}
R.~Chajwa, N.~Menon, S.~Ramaswamy, and R.~Govindarajan, ``Waves, algebraic
  growth, and clumping in sedimenting disk arrays,'' {\em Phys. Rev. X},
  vol.~10, p.~041016, Oct 2020.

\bibitem{IVERSEN2020102445}
M.~H. Iversen and R.~S. Lampitt, ``Size does not matter after all: No evidence
  for a size-sinking relationship for marine snow,'' {\em Progress in
  Oceanography}, vol.~189, p.~102445, 2020.

\bibitem{Martin_1987}
J.~H. Martin, G.~A. Knauer, D.~M. Karl, and W.~W. Broenkow, ``{VERTEX}: carbon
  cycling in the northeast pacific,'' {\em Deep Sea Research Part A.
  Oceanographic Research Papers}, vol.~34, pp.~267--285, feb 1987.

\bibitem{Lauderdale2021}
J.~M. Lauderdale and B.~B. Cael, ``Impact of remineralization profile shape on
  the air-sea carbon balance,'' {\em Geophysical Research Letters}, vol.~48,
  Apr. 2021.

\bibitem{Bidle2015}
K.~D. Bidle, ``The molecular ecophysiology of programmed cell death in marine
  phytoplankton,'' {\em Annual Review of Marine Science}, vol.~7, no.~1,
  pp.~341--375, 2015.

\bibitem{Alldredge1988}
A.~L. Alldredge and C.~Gotschalk, ``In situ settling behavior of marine
  snow1,'' {\em Limnology and Oceanography}, vol.~33, pp.~339--351, May 1988.

\bibitem{Picheral2010}
M.~Picheral, L.~Guidi, L.~Stemmann, D.~M. Karl, G.~Iddaoud, and G.~Gorsky,
  ``The underwater vision profiler 5: An advanced instrument for high spatial
  resolution studies of particle size spectra and zooplankton,'' {\em Limnology
  and Oceanography: Methods}, vol.~8, no.~9, pp.~462--473, 2010.

\bibitem{Trudnowska2021}
E.~T. et~al., ``Marine snow morphology illuminates the evolution of
  phytoplankton blooms and determines their subsequent vertical export,'' {\em
  Nature Communications}, vol.~12, May 2021.

\bibitem{Giering2022}
J.~R. Williams and S.~L.~C. Giering, ``In situ particle measurements
  deemphasize the role of size in governing the sinking velocity of marine
  particles,'' {\em Geophysical Research Letters}, vol.~49, no.~21,
  p.~e2022GL099563, 2022.
\newblock e2022GL099563 2022GL099563.

\bibitem{Krishnamurthy_2020}
D.~Krishnamurthy, H.~Li, F.~B. du~Rey, P.~Cambournac, A.~G. Larson, E.~Li, and
  M.~Prakash, ``Scale-free vertical tracking microscopy,'' {\em Nature
  Methods}, vol.~17, pp.~1040--1051, aug 2020.

\bibitem{Park2018}
Y.~Park, C.~Depeursinge, and G.~Popescu, ``Quantitative phase imaging in
  biomedicine,'' {\em Nature Photonics}, vol.~12, pp.~578--589, Sept. 2018.

\bibitem{Witten_2020}
T.~A. Witten and H.~Diamant, ``A review of shaped colloidal particles in
  fluids: anisotropy and chirality,'' {\em Reports on Progress in Physics},
  vol.~83, p.~116601, oct 2020.

\bibitem{10.3389/fmars.2020.00564}
S.~L.~C. Giering, B.~Hosking, N.~Briggs, and M.~H. Iversen, ``The
  interpretation of particle size, shape, and carbon flux of marine particle
  images is strongly affected by the choice of particle detection algorithm,''
  {\em Frontiers in Marine Science}, vol.~7, 2020.

\bibitem{McDonnell_2012}
A.~M.~P. McDonnell and K.~O. Buesseler, ``A new method for the estimation of
  sinking particle fluxes from measurements of the particle size distribution,
  average sinking velocity, and carbon content,'' {\em Limnology and
  Oceanography: Methods}, vol.~10, pp.~329--346, may 2012.

\bibitem{MCCAVE1975491}
I.~McCave, ``Vertical flux of particles in the ocean,'' {\em Deep Sea Research
  and Oceanographic Abstracts}, vol.~22, no.~7, pp.~491--502, 1975.

\bibitem{PASSOW2002287}
U.~Passow, ``Transparent exopolymer particles (tep) in aquatic environments,''
  {\em Progress in Oceanography}, vol.~55, no.~3, pp.~287--333, 2002.

\bibitem{McNown1950}
J.~S. McNown and J.~Malaika, ``Effects of particle shape on settling velocity
  at low reynolds numbers,'' {\em Transactions, American Geophysical Union},
  vol.~31, no.~1, p.~74, 1950.

\bibitem{nissanka_ma_burton_2023}
K.~Nissanka, X.~Ma, and J.~C. Burton, ``Dynamics of mass polar spheroids during
  sedimentation,'' {\em Journal of Fluid Mechanics}, vol.~956, p.~A28, 2023.

\bibitem{Christensen1984}
R.~M. Christensen and L.~B. Freund, ``Theory of viscoelasticity, an
  introduction, 2nd edition,'' {\em J. Appl. Mech.}, vol.~51, pp.~226--226,
  Mar. 1984.

\bibitem{Happel1983}
J.~Happel and H.~Brenner, {\em Low Reynolds number hydrodynamics}.
\newblock Springer Netherlands, 1983.

\bibitem{KIM1991}
S.~Kim and S.~J. Karrila, {\em Microhydrodynamics}.
\newblock Elsevier, 1991.

\bibitem{Guo2021}
C.~Guo, J.~Sun, X.~Wang, S.~Jian, M.~Abu~Noman, K.~Huang, and G.~Zhang,
  ``Distribution and settling regime of transparent exopolymer particles (tep)
  potentially associated with bio-physical processes in the eastern indian
  ocean,'' {\em Journal of Geophysical Research: Biogeosciences}, vol.~126,
  no.~4, p.~e2020JG005934, 2021.
\newblock e2020JG005934 2020JG005934.

\bibitem{doi:10.1146/annurev-fluid-010719-060107}
M.~Alves, P.~Oliveira, and F.~Pinho, ``Numerical methods for viscoelastic fluid
  flows,'' {\em Annual Review of Fluid Mechanics}, vol.~53, no.~1,
  pp.~509--541, 2021.

\bibitem{Meng2016}
S.~Meng and Y.~Liu, ``New insights into transparent exopolymer particles
  ({TEP}) formation from precursor materials,'' {\em Scientific Reports},
  vol.~6, Jan. 2016.

\bibitem{Mrokowska2022}
M.~M.~M. et~al., ``Effect of exopolymer gels on the viscoelasticity of
  mucus-rich saltwater and settling dynamics of particles,'' {\em Marine
  Chemistry}, vol.~246, p.~104163, Oct. 2022.

\bibitem{Jenkinson1986}
I.~R. Jenkinson, ``Oceanographic implications of non-newtonian properties found
  in phytoplankton cultures,'' {\em Nature}, vol.~323, pp.~435--437, Oct. 1986.

\bibitem{Laber2018}
C.~P.~L. et~al., ``Coccolithovirus facilitation of carbon export in the north
  atlantic,'' {\em Nature Microbiology}, vol.~3, pp.~537--547, Mar. 2018.

\bibitem{Mari1999}
X.~Mari, ``Carbon content and c:n ratio of transparent exopolymeric particles
  ({TEP}) produced by bubbling exudates of diatoms,'' {\em Marine Ecology
  Progress Series}, vol.~183, pp.~59--71, 1999.

\bibitem{Engel2002}
A.~Engel, ``Direct relationship between {CO}2 uptake and transparent exopolymer
  particles production in natural phytoplankton,'' {\em Journal of Plankton
  Research}, vol.~24, pp.~49--53, Jan. 2002.

\bibitem{Arrigo2007}
K.~R. Arrigo, ``Marine manipulations,'' {\em Nature}, vol.~450, pp.~491--492,
  Nov. 2007.

\bibitem{10.1093/pnasnexus/pgac311}
B.~Borer, I.~H. Zhang, A.~E. Baker, G.~A. O'Toole, and A.~R. Babbin, ``{Porous
  marine snow differentially benefits chemotactic, motile, and nonmotile
  bacteria},'' {\em PNAS Nexus}, vol.~2, p.~pgac311, 12 2022.

\bibitem{Ben2012}
A.~Vardi, L.~Haramaty, B.~A. S.~V. Mooy, H.~F. Fredricks, S.~A. Kimmance,
  A.~Larsen, and K.~D. Bidle, ``Host–virus dynamics and subcellular controls
  of cell fate in a natural coccolithophore population,'' {\em Proceedings of
  the National Academy of Sciences}, vol.~109, no.~47, pp.~19327--19332, 2012.

\bibitem{Erickson2009}
H.~P. Erickson, ``Size and shape of protein molecules at the nanometer level
  determined by sedimentation, gel filtration, and electron microscopy,'' {\em
  Biological Procedures Online}, vol.~11, pp.~32--51, May 2009.

\bibitem{DUCKLOW2001217}
D.~A. Hansell and H.~W. Ducklow, ``Bacterioplankton distribution and production
  in the bathypelagic ocean: Directly coupled to particulate organic carbon
  export?,'' {\em Limnology and Oceanography}, vol.~48, no.~1, pp.~150--156,
  2003.

\bibitem{Lee2022}
N.~Haëntjens, E.~S. Boss, J.~R. Graff, A.~P. Chase, and L.~Karp-Boss,
  ``Phytoplankton size distributions in the western north atlantic and their
  seasonal variability,'' {\em Limnology and Oceanography}, vol.~67, no.~8,
  pp.~1865--1878, 2022.

\bibitem{Lupp2003}
C.~Lupp, M.~Urbanowski, E.~P. Greenberg, and E.~G. Ruby, ``The vibrio fischeri
  quorum-sensing systems ain and lux sequentially induce luminescence gene
  expression and are important for persistence in the squid host,'' {\em
  Molecular Microbiology}, vol.~50, no.~1, pp.~319--331, 2003.

\bibitem{doi:10.1073/pnas.95.12.6578}
W.~B. Whitman, D.~C. Coleman, and W.~J. Wiebe, ``Prokaryotes: The unseen
  majority,'' {\em Proceedings of the National Academy of Sciences}, vol.~95,
  no.~12, pp.~6578--6583, 1998.

\bibitem{10.3389/fpls.2018.00869}
H.~Wang, R.~Zhu, J.~Zhang, L.~Ni, H.~Shen, and P.~Xie, ``A novel and convenient
  method for early warning of algal cell density by chlorophyll fluorescence
  parameters and its application in a highland lake,'' {\em Frontiers in Plant
  Science}, vol.~9, 2018.

\bibitem{doi:10.1146/annurev-marine-010213-135138}
E.~A. D'Asaro, ``Turbulence in the upper-ocean mixed layer,'' {\em Annual
  Review of Marine Science}, vol.~6, no.~1, pp.~101--115, 2014.

\bibitem{doi:10.1073/pnas.1602307113}
K.~Son, F.~Menolascina, and R.~Stocker, ``Speed-dependent chemotactic precision
  in marine bacteria,'' {\em Proceedings of the National Academy of Sciences},
  vol.~113, no.~31, pp.~8624--8629, 2016.

\bibitem{Milo2015}
R.~Milo and R.~Phillips, {\em Cell Biology by the Numbers}.
\newblock Garland Science, Dec. 2015.

\bibitem{doi:10.1146/annurev-marine-122414-033938}
D.~L. Kirchman, ``Growth rates of microbes in the oceans,'' {\em Annual Review
  of Marine Science}, vol.~8, no.~1, pp.~285--309, 2016.
\newblock PMID: 26195108.

\bibitem{aa6a7d8b-1f86-39e2-ad70-6d83b2b27ded}
R.~S. et~al., ``Microbial decomposition of large organic particles in the
  northwestern mediterranean sea: an experimental approach,'' {\em Marine
  Ecology Progress Series}, vol.~198, pp.~61--72, 2000.

\end{thebibliography}
\bibliographystyle{ieeetr} 


\pagebreak
\clearpage
\newpage
\onecolumngrid
\section*{Supplementary Material for Hidden comet-Tails of Marine Snow}
\section{Sample Collection}
Samples were collected aboard the R/V Endeavor, as part of the Research at the Interface of Phytoplankton, Particle, lipids and Export (RIPPLE) project cruise (11-23 June), 2021 in the Gulf of Maine. Freely hanging sediment traps in Lagrangian mode \cite{Ben2015} were deployed at depths of 50, 80 or 150 meters for 24 hours before recovery. The data presented in this work is analysed from the following trap deployments at 80m depth:
 \begin{center}
\begin{tabular}{ |p{2.5cm}| p{2.5cm}|p{2.5cm}|p{2.5cm}| p{2.5cm} | p{2.5cm} |  }
 \hline
 \multicolumn{5}{|c|}{\textbf{RIPPLE Net trap sampling}} \\
 \hline
 Date& Time deployed & Date & Time recovered & Depth (m)\\
 \hline
  6/15/21 & 0935    & 6/16/21 & 0930 & 80 \\
  6/16/21 & 0951    & 6/17/21 & 0850 & 80 \\
  6/17/21 & 0919    & 6/18/21 & 0945 & 80 \\
  6/18/21 & 1015    & 6/19/21 & 1107 & 80 \\
  6/19/21 & 0952    & 6/20/21 & 0843 & 80 \\
  6/20/21 & 2119    & 6/21/21 & 1020 & 80 \\
 \hline
\end{tabular}
\end{center}
\onecolumngrid
\begin{figure*}[b]
      \begin{center}
      \includegraphics[width=14.5 cm]{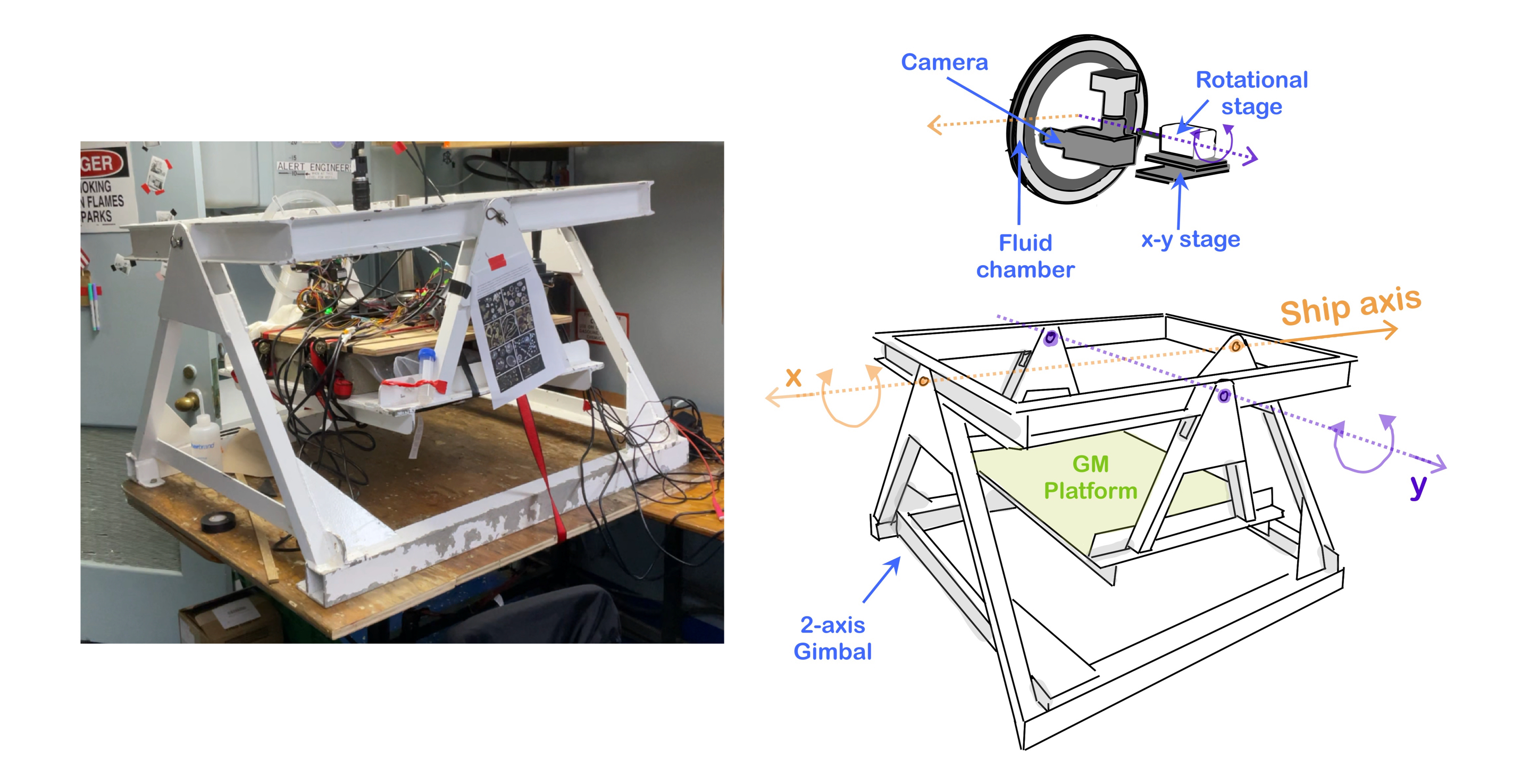}
      \caption{\label{Fig1S} \textbf{Gravity machine setup with 2-axis gimbal}
      }
     \end{center}
\end{figure*}
In this study, we focus on the response of the organic debris (marine snow) to the hydrodynamic stresses generated by gravitational forcing. The fluid dynamical and mechanical perturbations were minimized while handling the sample to preserve its structure (being on a ship); however, our approach applies to the aggregates that are made up of the ``raw material" collected in the sediment trap. With less emphasis on the preservation of structure, we studied the material response. We acknowledge that the aggregates that are present in the sediment trap are not a true reflection of the shape and size of the sinking aggregate, an ongoing technical limitation in the field for sampling marine snow using sediment trap. However, sediment traps does offer us the true material building blocks of marine snow and carbon flux. And we show that it is possible to systematically get useful physical quantities that are directly related to the carbon flux.
\section{Data collection}
Data was collected while on board the R/V Endeavor during the cruise using a gravity machine \cite{Krishnamurthy_2020}. We mix 60 $\mu L$ of $700 nm - 2 \mu m$ polystyrene bead solution in 100 $mL$ Sea Water for doing Particle Imagining Velocimetry (PIV). We load about 1mL of the sediment from the collected sample in the gravity machine wheel containing the bead sea water solution. After loading the aggregates, the marine snow suspension was gently homogenized by manually rotating the wheel clockwise and counter-clockwise multiple times. This homogenization is needed to minimize inter-particle hydrodynamic interactions. We then take tracks of marine snow aggregates for around 2 min at 5fps. 
\section{Analysis Pipeline}
\subsection{Effective radius and n\"{a}ive Stokesian settling approximation}
Gravity machine directly gives the virtual vertical distance traversed as a function of time $z(t)$. The effective vertical position as a function of time, shows fluctuation in vertical velocity due to the ship motion, but this disturbance only adds a residual error in a linear fit, thanks to the gimbal. This gives us the vertical settling speed. To correlate it with the visible particle size we threshold the images after conversion to 8-bit and measure the projected area and from this get the effective radius of a circle with an equivalent area.
\subsection{Mucus halo measurement}
The collected images and tracks were initially processed in a customized ImageJ macros. The vibrations of the ship resulted in misalignment of aggregates in adjacent frames. To remove this noise we automated the image registration process using a Matlab script. A command-line based PIV was conducted on the resulting images, on Matlab PIVLab. The PIV data and GM tracking data were then coherently analysed for mucus and particulate matter quantification using a custom Matlab function. All the image processing and analysis were automated, the codes for which are made available. 

\section{Alcian Blue Staining}
Particles were stained with alcian blue (AB) to visualize TEP as per previously published protocols \cite{Ben2012}. A stock solution (1\% alcian blue in 3\% acetic acid in water) was used to prepare a working solution ( 0.04\% AB). The working solution was then passed through a 0.2 $\mu$m pore-size syringe filter prior to staining. Water samples containing marine snow `splits' from the net trap were gently filtered onto 0.4 $\mu$m pore-size polycarbonate filters using low and constant vacuum pressure (less than 200 mbar). Once filtered, 500 $\mu$L of the AB working solution was added and the pump was switched on immediately to remove the stain. The filter was then rinsed with 500 $\mu$ L of double distilled water under gentle vacuum. The filter was then placed in an eppendorf tube with 1.5 mL of 0.2$\mu$m filtered sea water and the sample was agitate to release marine snow particles off of the filter and into the bottom of the eppendorf tube. Samples were visualized immediately and kept at $-20$ Degree C for additional analyses.

\section{Sample Preservation}
Particles were preserved in a 4$\%$ Paraformaldehyde solution in Phosphate-Buffered Saline solution for 3D structural analysis. 

\section{Fluorescence Microscopy}
Fluorescence microscopy images were collected using a LSM780 Confocal Microscope through the Cell Sciences Imaging Facility at Stanford Univeristy. Samples were mounted in 35mm glass-bottom petri dishes, and were uncompressed. Samples were not stained. Any color present comes from autofluorescence of the particle. 

\section{Holotomography}
3D refractive index images were collected using holotomography on the Tomocube microscope. Samples were mounted in tomodishes and were compressed by 22mmx22mm coverslips. 

\section{Minimal model for the Comets of mucus}
\begin{figure}[h]
      \begin{center}
      \includegraphics[width=12 cm]{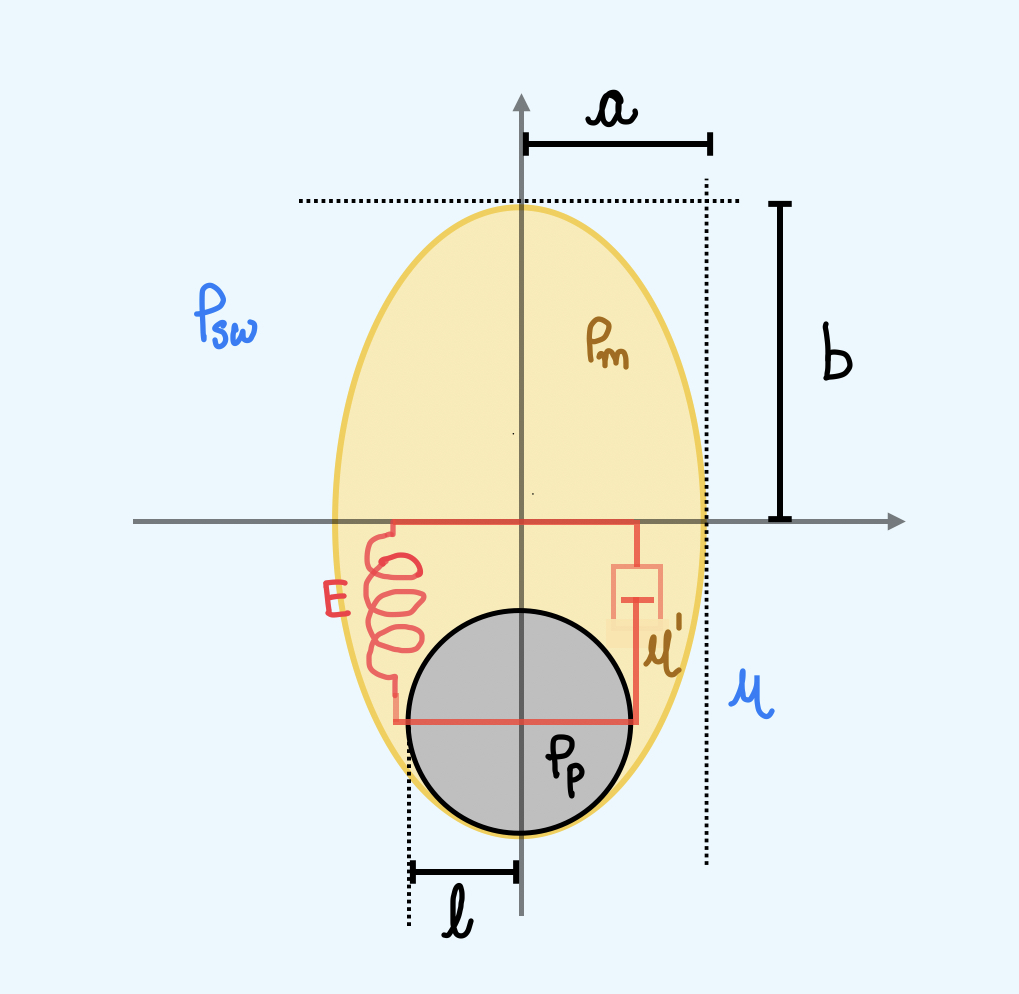}
      \caption{\label{Fig2S} \textbf{Sinking with viscoelastic mucus halo} 
      }
     \end{center}
\end{figure} 
\subsection{Hydrodynamic Stretching of Tails}
To account for the deformation of the mucus halo, we minimally approximate the mucus blob as a viscoelastic spring subjected to the stress field due to sedimenting particulate aggregate attached to the halo. The sedimentation self-stress of a force monopole  located at the origin, along gravity axis $-\hat{z}$ is
\begin{align}
    \sigma_{ij} & = \mu ( \partial_{j} v_{i} + \partial_{i} v_{j} )\\
    & = \frac{F^{k}}{8} \left[ -\frac{1}{r^3} (\delta_{ik} r_j + \delta_{jk} r_i) + \frac{1}{r^3} (2 \delta_{ij} r_k + \delta_{kj} r_{i} + \delta_{ki}r_{j}) - \frac{6}{r^5} r_i r_j r_k \right]
\end{align}
Here, the strength of the monopole $F^k$ incorporates the effective buoyancy of particulate matter and mucus combined
\begin{equation}
    \vect{F} = V (\rho_{sw} - \rho_{m})\left[\frac{V_{p}}{V} \left(\frac{\rho_{p} - \rho_{m}}{\rho_{sw} - \rho_{m}}\right) -1 \right] \vect{g} 
\label{effective_force}
\end{equation}
Particulate matter, initially embedded in the bulk of mucus, being heavier than the surrounding mucus falls to the boundary of the mucus domain. We assume that the sedimentation time-scales $T = 9 \mu/ 2 g (\rho_{p} - \rho_{m}) l$ associated with this fall is much smaller than the mucus deformation time-scale $ \tau = \mu'/E$, where $\mu'$ and $E$ are the mucus viscosity and the Young's modulus respectively; thus it is assumed that the aggregate is present at the boundary of the mucus at all times. We treat the mucus relative extension $\delta \vect{b} /b_{0} \equiv (b/b_{0} - 1)  \hat{z}$ as a visco-elastic spring evolving via the Kelvin-Voigt model
\begin{equation}
     \dot{\delta b_{i}} + \ \frac{E}{\mu'} \delta b_{i} = \frac{1}{\mu'} \, b_{0} \, \sigma_{ij}\delta b_{j}
     \label{KV}
\end{equation}
When the extensional axis is aligned with gravity axis, thanks to the polarity of the marine snow, combining (2) and \eqref{effective_force} gives the self-stress 
\begin{equation}
    \vect{\sigma} \cdot \vect{b} = \frac{1}{2 \pi} {V (\rho_{sw} - \rho_{m})g} \left( \frac{V_p}{V} \tilde{\rho} - 1\right) \frac{1}{b^2 (1 - l/b)^2}\hat{z}
    \label{self_stress}
\end{equation}
where $\tilde{\rho} = {(\rho_{p} - \rho_{m})}/{(\rho_{sw} - \rho_{m})}$. Expanding the denominator in \eqref{self_stress} as a power series in $l/b_{0}$ gives
\begin{equation}
     \frac{1}{b^2 (1 - l/b)^2} =  \frac{1}{b_{0}^2(1 + \delta b)^2} \left\{ 1 - \frac{2}{1 + \delta b} \left(\frac{l}{b_{0}} \right) + \mathcal{O}\left(\frac{l^2}{b_0^2}\right) + ...  \right\}
\end{equation}

For non-zero mucus volume $l/b$ is always less than 1, thus we retain only the leading term in $l/b_{0}$, which gives the sedimentation self-stress
\begin{equation}
    \vect{\sigma} \cdot \vect{b} = \frac{2}{3} \frac{a_{0} (\rho_{sw} - \rho_{m})g}{(1 + \delta b/ a_{0})^2} \left( \frac{V_p}{V} \tilde{\rho} - 1\right)
    \label{reduced_selfstress}
\end{equation}

Note that we have imposed volume conservation and specifed initial condition $ b_0 = a_{0} > l$ i.e. initial particle shape is isotropic and there is a non-zero mucus volume. Combining \eqref{KV} and \eqref{reduced_selfstress} and non-dimensionalizing using length scale $L= a_{0}$ and sedimentation time scale $T = 9 \mu/ 2 g (\rho_{sw} - \rho_{m}) a_{0}$ gives the equation for deformation of the viscoelastic spring

\begin{equation}
    \dot{\delta b } + \underbrace{\frac{\mu}{\mu'} \, \mathcal{D} \, \delta b}_{\text{Visco-elastic relaxation}} = \underbrace{3 \frac{\mu}{\mu'}  \left[\frac{V_{p}}{V} \tilde{\rho} - 1 \right] \frac{1}{(1 + \delta b)^2}}_{\text{Sedimentation self-stress}}
    \label{KV_nondimensional}
\end{equation}
where $\mathcal{D} \equiv 9E/[2g(\rho_{p}-\rho_{m})a_{0} ]$ is the ratio of elastic to gravitational response. At long times \eqref{KV_nondimensional} yields two asymptotic scaling regimes for very small and very large deformations
\begin{equation}
  \delta b = \begin{cases}
               3 \left(\frac{V_{p}}{V} \tilde{\rho} - 1\right) \mathcal{D} \,\,\,\, \text{for} \,\,\delta b \ll 1\\
              \left[3\left(\frac{V_{p}}{V}\tilde{\rho} - 1\right)\right]^{1/3} \mathcal{D}^{-1/3} \, \,\,\, \text{for} \,\,\delta b \gg 1 
            \end{cases}
            \label{tail_scaling}
\end{equation}
Note that the viscosity sets the time scale to achieve the steady state, however, steady state is independent of the viscosity $\mu$ and $\mu'$. We now turn to the distortion of the mucus in the degenerate transverse direction. The volume conservation of the mucus $V_{m}$, in the prolate spheroid approximation gives $a$, which does not have an independent dynamics
\begin{equation}
    \frac{a}{l} = \left[ \frac{V}{V_{p}}\frac{l}{b}\right]^{1/2}
    \label{mucus_conservation}
\end{equation}
which gives dynamic eccentricity of the mucus halo $e$ as a function of tail extension $\delta b$

\begin{equation}
    e = \left[ 1 - (1+\delta b)^{-3} \right]^{1/2}
\end{equation}



\subsection{Sedimentation with mucus halo}

In the fluid-mechanical regime where viscous forces dominate over the inertial forces, the sinking speed for a sphere of radius $l$, density $\rho$ in an ambient fluid of viscosity $\mu$ and density $\rho_{sw}$ is given by the celebrated Stokes law
\begin{equation}
    \frac{d \vect{z}}{dt} = \frac{2 \vect{g} (\rho_{eff} - \rho_{sw}) \, l^2 }{9 \mu}
    \label{Stokes}
\end{equation}
Here, $\vect{g}$ is the acceleration due to gravity. In our minimal model we assume that the mucus halo to be ellipsoidal with semi-major and semi-minor axes $b$ and $a$ respectively and density $\rho_{m}$. The visible particulate aggregate with effective radius $l$ has a density $\rho_p$ greater than $\rho_{m}$.

In the fluid-mechanical regime where viscous forces dominate over the inertial forces, the sinking speed spheroid with semi-minor axis $a$ and semi-major axis $b$, density $\rho$ in an ambient fluid of viscosity $\mu$ and density $\rho_{sw}$ is given by the mobility relation
\begin{equation}
    U_{i} = \frac{F^{j}}{6\pi \mu b} \left[ X_{A}^{-1} d_{i}d_{j} + Y_{A}^{-1} (\delta_{ij} - d_{i}d_{j}) \right], 
\end{equation}

using the Einstein summation convention, where $\vect{d}$ is the unit orientation vector along the symmetry axis of the particle, and $X_{A}^{-1}$ and $Y_{A}^{-1}$ are the mobility functions that depends on the eccentricity, $e = \sqrt{1- (a/b)^2}$, which for prolate spheroids take the form \cite{KIM1991}:
\begin{align}
X_{A}^{-1} & = \frac{3}{8 e^3}\left[ (1 + e^2)\ln \left( \frac{1 + e}{1 - e}\right) - 2 e\right ] \nonumber  \\
Y_{A}^{-1} & = \frac{3}{16 e^3}\left[ (3 e^2 - 1)\ln \left( \frac{1 + e}{1 - e}\right) + 2 e\right ] \nonumber
\end{align}

These account for the shape dependent sinking of the aggregate. Since, the visible aggregates are generally heavier than the mucus halo, an effective particle (particle + mucus) can be treated as a bottom heavy particle that aligns with the gravity axis $\vect{g}$ in a steady state ($\vect{d} \times \vect{g} = 0 $). Therefore for the present analysis only $X_{A}^{-1}$ will contribute. Note that the external force $F^{j}$ is the buoyant weight of the particle, given by

\begin{equation}
    \vect{F} = V (\rho_{eff} - \rho_{sw}) \vect{g}
\end{equation}
where, $V = 4 \pi a^2 b/3$ is the volume of the spheroid, $\vect{g}$ is the acceleration due to gravity. Defining the effective density of the marine snow with particulate matter and mucus combined, $\rho_{eff} = (\rho_p \, V_{p} \, + \, \rho_{m} V_{m} )/V$, where $V_{p}$, $V_{m}$ and $V$ is the volume of the particle, mucus and the total volume respectively (note that $ V = V_{p} + V_{m} $). 

\begin{align}
    \frac{dz}{dt} & = \frac{V g}{ 6 \pi \mu b} \left( \rho_p \frac{V_{p}}{V} \, + \, \rho_m \frac{V_{m}}{V}  \, - \, \rho_{sw} \right) X_{A}^{-1}\\
    & = \frac{g}{6 \pi \mu b} \left[ \rho_p V_p + \left( \frac{4}{3} \pi a^2 b - V_p \right) \rho_m + \frac{4}{3} \pi a^2 b \rho_{sw} \right] X_{A}^{-1}
\end{align}


Non-dimensionalizing using the characteristic length scale $L = a_{0}$ and characteristic time scale $T = 9 \mu/ 2 g (\rho_{sw} - \rho_{m}) a_{0}$ gives

\begin{equation}
\frac{d \tilde{z}}{dt'} = \left( \frac{V_{p}}{V} \tilde{\rho} - 1 \right) \frac{X_{A}^{-1}}{1 + \delta b}
\label{ND_sinking}
\end{equation}
where $\tilde{z} \equiv z/a_{0}$. Substituting the form of mobility function $X_{A}^{-1}$  for prolate spheroid gives the scaling relation for the non-dimensional sinking speed $\tilde{U}_z$,
\begin{equation}
    \tilde{U}_{z} \, =  f\left( e, \frac{\rho_p - \rho_m}{\rho_{sw} - \rho_m} \right)
\end{equation}
where the scaling function is
\begin{equation}
   f\left( e , \frac{\rho_p - \rho_m}{\rho_{sw} - \rho_m} \right)  \equiv \frac{3}{8 e^3}\left[ (1 + e^2)\ln \left( \frac{1 + e}{1 - e}\right) - 2 e\right ] \times \left( \frac{V_{p}}{V} \tilde{\rho} - 1 \right) \frac{1}{ 1 + \delta b}.
\end{equation} 

Note that the eccentricity of the spheroid $e$ is a function of $a/b$; the first half of the scaling function is purely geometrical while the second half incorporates the physical parameters (relevant densities).  In the above scaling, we have not accounted for the dependence of drag on the shape of the emergent particle, that could be elongated. 

\section*{Sedimentation based Rheology of mucus}
\begin{figure}[h]
      \begin{center}
      \includegraphics[width=18 cm]{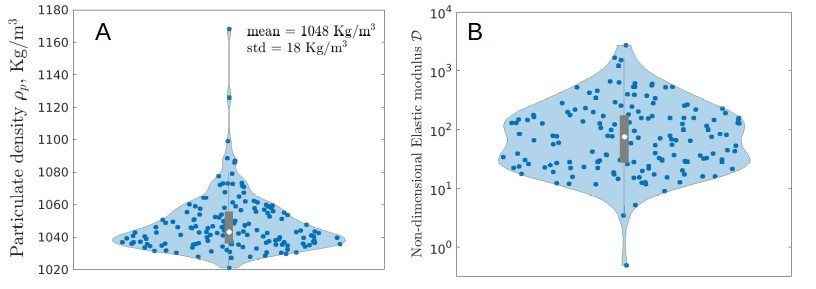}
      \hspace{0cm} \newline 
      \caption{\label{Fig3S} \textbf{Sedimentation based Rheology:} A) The density distribution of the visible aggregate after accounting for the mucus [equation \eqref{particulate_density}]. B) 
      }
     \end{center}
\end{figure} 
In our simple framework, the linear viscoelastic properties of mucus is lumped into a dynamic loss modulus $G''$ and a storage modulus $G'$ which gives the effective viscosity and elastic modulus $E$ respectively. Mucus stretches in response to the sedimentation stress. In the steady state approximation of \eqref{KV_nondimensional} it is possible to get a rough estimate of the elastic modulus $E$ of the mucus from our measurement, which is the zero frequency limit of the storage modulus $ \lim_{\omega \to 0} G'$, for sedimentation stresses are nearly constant (thanks to the gimbal).

We directly measure the sinking speed $U_{z}$, shape dependent mobility $X_{A}^{-1}$, mucus distortion $\delta b$ and relative volumes $V_{p}/V$. Using equation \eqref{ND_sinking} we get the non-dimensional density
\begin{equation}
    \tilde{\rho} =  \frac{V}{V_{p}}\left[ 1 + \frac{\tilde{U_{z}} (1 + \delta b)}{X_{A}^{-1}} \right] 
    \label{ND_rho}
\end{equation}
Here calculating non-dimensional sinking speed $\tilde{U_{z}}$ from $U_{z}$ requires knowing $\rho_{sw} - \rho_{m}$. The density of mucus (like the visco-elasticity) in the open ocean depends on a multitude of biochemical factors and is poorly understood. We use the value $ \rho_{sw} - \rho_{m} = 0.5 Kg/m^3$ based on a recent coarse-grained observation of the sinking speed \cite{Guo2021}. Mucus being slightly lighter than the sea water allows for positive buoyancy which is observed in TEP. From equation $\eqref{ND_rho}$ we get the density of the visible portion of the marine snow

\begin{equation}
    \rho_{p} = (\rho_{sw} - \rho_{m}) \tilde{\rho} + \rho_{m}
    \label{particulate_density}
\end{equation}
Substituting $\tilde{\rho}$ in the steady state limit of the comet tail equation \eqref{KV_nondimensional} gives the storage modulus of the visco-elastic mucus
\begin{equation}
   \lim_{\omega \to 0} G' = \frac{3\mu}{T}\left(\frac{V_{p}}{V} \tilde{\rho} -1 \right)\frac{1}{\delta b (1 + \delta b)^2}
    \label{rheology}
\end{equation}

\section{Residence time Comparison}
Qualitatively the presence of mucus reduces the effective density of the marine snow and makes the aggregates falls slower than without mucus. We quantify this discrepancy between measurement by calculating the residence time of the aggregate to cross the 100 m depth in the ocean, 1) without accounting for the mucus and using the class Stokes law and 2) accounting for modified effective density and size spectrum due to the mucus. 

1) The vertical speed estimated from the naive stokes law is $\dot{U_{z}} = \frac{2 g}{9 \mu} (\bar{\rho_{p}} - \rho_{sw}) \ell$, where $\bar{\rho_{p}} = 1048 Kg/m^{3}$ is the mean particulate density measured from the sedimentation based rheology [see Fig. 3 A) above], and $\ell$ is the equivalent radius of the measured visible size spectra [see Figure 3E in main text]. This gives the residence time in the upper $100$m in the ocean in seconds 
\begin{equation}
    \tau_{100} = \frac{100 \mu}{\frac{2}{9} g (\bar{\rho} - \rho_{sw}) \ell^2}
\end{equation}

2) Accounting for mucus gives two effective variables that can be measured, extent of mucus by volume  $V_{p}/V$ and extension of the halo $\delta b$, assuming a fixed particulate density $\bar{\rho_{p}} = 1048 Kg/m^{3}$ and mucus density $\rho_{m} = \rho_{s} - 0.5 Kg/m^{3}$ which is taken to be slightly smaller than sea water density $\rho_{s} = 1025 Kg/m^{3}$ to allow for the possibility of slowly rising TEP particles, as observed in \cite{Guo2021}. This fixes the non-dimensional density parameter $\tilde{rho} = \rho_{p} - \rho_{m}/(\rho_{sw} - \rho_{m})$. Note that, there is a lack of direct measurement of density and visco-elastic properties of marine mucus. Using equation 5 in main text the residence time in the upper 100 m becomes
\begin{equation}
    \tau'_{100} = \frac{100 (1 + \delta b) \mu}{\frac{2}{9} g (\rho_{sw} - \rho_{m}) {a_{0}}^2 (V_{p} \tilde{\rho}/V -1) \mathbb{X}^{-1}(e)}
\end{equation}
where the hydrodynamic mobility $\mathbb{X}^{-1}(e)$ depends on the eccentricity $e$ of the fitted ellipse around the two-phase marine snow particle and is estimated by direct measurement. 

\vspace{0.5cm}
The measured distribution of $\tau_{100}$ and $\tau'_{100}$ is plotted in Figure 4 F.

\section{Multiple Length and Time Scales in Biological Pump}
Tables below shows the non-exhaustive list of typical length and time scales involved in the Biological pump and the source of the data 
 \begin{center}
\begin{tabular}{ |p{5cm}| p{3cm}| p{7cm} |  }
 \hline
 \multicolumn{3}{|c|}{\textbf{Multiple length scales}} \\
 \hline
  & Length scale (m) & Source\\
 \hline
 Biomolecules & $10^{-8}$ & \cite{Erickson2009}{H. P. Erickson et al., 2009 }\\
 Marine Bacteria & $10^{-6}$ & \cite{DUCKLOW2001217} {H. Ducklow, 2003} \\
 Phytoplankton & $10^{-5} - 10^{-4}$ & \cite{Lee2022} {N. H\"{a}entjen et al., 2022}\\
 Quorum sensing distance & $10^{-5}$ & \cite{Lupp2003} {C. Lupp, et al. 2003}\\
 Marine bacteria separation & $10^{-4}$ & \cite{doi:10.1073/pnas.95.12.6578}{W. B. Whitman et al., 1998.}\\
 Marine snow & $10^{-4}>$& \cite{LAMPITT2001160} {R. Lampitt, 2001}\\
 Phytoplankton separation& $10^{-3}$ & \cite{10.3389/fpls.2018.00869} {H. Wang et al., 2018}\\
 Turbulent boundary layer & $10^{1} - 10^{2}$ & \cite{doi:10.1146/annurev-marine-010213-135138} {E. A. D'Asaro, 2014}\\
 Euphotic zone & $10^{2}$ & \href{https://www.noaa.gov/jetstream/ocean/layers-of-ocean#:~:text=This%20surface%20layer%20is%20also,of%20the%20visible%20light%20exists.}{NOAA}\\
 Ocean Floor & $10^{4}$ & \href{https://www.noaa.gov/jetstream/ocean/layers-of-ocean#:~:text=This%20surface%20layer%20is%20also,of%20the%20visible%20light%20exists.}{NOAA}\\
 \hline
\end{tabular}
\end{center}
The bacterial and plankton separation are calculated by taking the cube root of the known inverse number density in freely suspended phase. For quorum sensing we use the typical critical density to calculate the length scale \cite{Lupp2003}.

 \begin{center}
\begin{tabular}{ |p{5cm}| p{3cm}| p{7cm} |  }
 \hline
 \multicolumn{3}{|c|}{\textbf{Multiple time scales}} \\
 \hline
  & Time scale (min) & Source\\
 \hline
  Phytoplankton sedimentation & $10^{-3}$ &  \cite{Larson2022.08.19.504465} A. G. Larson et al., 2022 \\
 Bacterial swimming & $10^{-2}$ & \cite{doi:10.1073/pnas.1602307113} K. Son et al., 2016\\
  Molecular diffusion & 1 & \cite{Milo2015} R. Milo and R. Phillips, 2015 \\
  Marine snow Sedimentation & $10^{-3} - 10^{1}$ &  \cite{LAMPITT2001160} {R. Lampitt, 2001}\\
 Phytoplankton growth & $10^{3}$ & \cite{doi:10.1146/annurev-marine-122414-033938} D. L. Kirchman, 2016 \\
  Bacterial growth & $10^{4}$ & \cite{doi:10.1146/annurev-marine-122414-033938} D. L. Kirchman, 2016\\
 Microbial decomposition & $10^{3}-10^{5}$ & \cite{aa6a7d8b-1f86-39e2-ad70-6d83b2b27ded} {R. Semp\'{e}r\'{e} et al. 2000}  \\
 \hline
\end{tabular}
\end{center}

We construct the time scale associated with diffusion across a plankton of size $100 \mu m$ \cite{Milo2015}, using the Stokes Einstein relation for the typical size of bio-molecule in an aqueous medium of viscosity $\mu = 10^{-3} Kg m^{-1} s^{-1}$. Sedimentation time-scale using Stokes law for a Phytoplankton of radius $100 \mu m$ and density $\rho = 1045 Kg/m^{3}$ which is a typical density of a cell \cite{Larson2022.08.19.504465}. The sedimentation time-scale of marine snow is defined as the time it takes for marine snow aggregate to sink by it own length. We refer to the data presented in \cite{LAMPITT2001160} to get the rough estimate.  We construct bacterial swimming time-scale from \cite{doi:10.1073/pnas.1602307113} by using the size-scale of the $100 \mu m$ phytoplankton.

\end{document}